\documentclass[fleqn,usenatbib,useAMS]{mnras}

\usepackage{newtxtext,newtxmath}
\usepackage[T1]{fontenc}
\usepackage{ae,aecompl}

\usepackage{graphicx}	
\usepackage{amsmath}	

\newcommand{\EqRef}[1]  {Eq.~(\ref{#1})}
\newcommand{\qufig}[1] {Fig.~\ref{#1}}
\newcommand{\element}[1] {{\ensuremath{\mathrm{#1}}}}
\newcommand{\D}{\mathrm{d}}
\newcommand{\E}{\mathrm{e}}
\newcommand{\T}{{\mathcal T}}
\newcommand{\cm}  {\ensuremath{\rm\, cm^{-1}}}
\newcommand{\mue} {\ensuremath{\rm\, \mu m}}

\title[Physical Retrieval of Exoplanet Temperatures]{Assessment of a Physics-based Retrieval of                     Exoplanet Atmospheric Temperatures from Infrared Emission Spectra}

\author[Franz Schreier et al.]{
Franz Schreier,$^{1}$\thanks{E-mail: franz.schreier@dlr.de (FS)}
J. Lee Grenfell,$^{2}$
Fabian Wunderlich,$^{2}$
and Thomas Trautmann$^{1}$
\\
$^{1}$DLR --- Deutsches Zentrum f\"ur Luft- und Raumfahrt, Institut f\"ur Methodik der Fernerkundung, 82234 Oberpfaffenhofen, Germany\\
$^{2}$DLR --- Deutsches Zentrum f\"ur Luft- und Raumfahrt, Institut f\"ur Planetenforschung, Rutherfordstr.\ 2, 12489 Berlin, Germany
}

\date{Accepted 2023 May 16. Received 2023 May 15; in original form 2023 February 22}

\pubyear{2023}

\begin{document}
\label{firstpage}
\pagerange{\pageref{firstpage}--\pageref{lastpage}}
\maketitle

\begin{abstract} 
Atmospheric temperatures are to be estimated from thermal emission spectra of Earth-like exoplanets orbiting M-stars as observed by current and future planned missions.
To this end, a line-by-line radiative transfer code is used to generate synthetic thermal infrared (TIR) observations.
The range of ``observed'' intensities provides a rough hint of the atmospheric temperature range without any a priori knowledge.
The equivalent brightness temperature (related to intensities by Planck's function) at certain wavenumbers can be used to estimate the atmospheric temperature at corresponding altitudes.
To exploit the full information provided by the measurement we generalize Chahine's original approach and infer atmospheric temperatures from all spectral data using the wavenumber-to-altitude mapping defined by the weighting functions.
Chahine relaxation allows an iterative refinement of this ``first guess''.
{Analysis of the $4.3\rm\,\mu m$ and $15\rm\,\mu m$ carbon dioxide TIR bands enables an estimate of atmospheric temperatures for rocky exoplanets even for low signal to noise ratios of 10 and medium resolution.
Inference of Trappist-1e temperatures is, however, more challenging especially for \element{CO_2} dominated atmospheres: the ``standard'' $4.3\rm\,\mu m$ and $15\rm\,\mu m$ regions are optically thick and an extension of the spectral range towards atmospheric window regions is important.
If atmospheric composition (essentially \element{CO_2} concentration) is known temperatures can be estimated remarkably well;
quality measures such as the residual norm provide hints on incorrect abundances.}
In conclusion, temperature in the mid atmosphere of Earth-like planets orbiting cooler stars can be quickly estimated from thermal IR emission spectra with moderate resolution.
\end{abstract}



\section{Introduction}

A quarter century after the detection of the first planet orbiting a main sequence star \citep{Mayor95} exoplanetary science has developed astonishingly quickly and is likely to continue on this path \citep{Merand21}.
Some 5000 exoplanets are known today\footnote{\url{http://exoplanet.eu/}}, several dedicated space missions are already in orbit for detection (Kepler/K2, Transiting Exoplanet Survey Satellite (TESS), \dots),
and characterisation (CHaracterising ExOPlanet Satellite \citep[CHEOPS,][]{Benz18}), and some others are in development, e.g.\ PLAnetary Transits and Oscillations of stars \citep[PLATO,][]{Rauer14etal}, Atmospheric Remote-sensing Infrared Exoplanet Large-survey \citep[ARIEL,][]{Tinetti18etal} or the proposed Large Interferometer For Exoplanets \citep[LIFE,][]{Defrere18,Quanz22e,Quanz22,Konrad22}.
The successfully launched James Webb Space Telescope (JWST) with its several infrared (IR) instruments will considerably expand our knowledge for a wide range of exoplanets \citep{Greene16}.

Atmospheric characterisation by means of microwave, IR, or ultraviolet spectroscopy is performed routinely for Earth as well as for the Solar System planets and moons \citep{Hanel03}
and its feasibility has also been demonstrated for extrasolar planets, mostly hot Jupiters \citep[e.g.][]{Madhusudhan09}.
For atmospheric remote sensing ``optimal estimation'' \citep{Rodgers76,Rodgers00} is by far the most common inversion technique
and this Bayesian method has also been used in an exoplanet context, e.g. \citet{Irwin08,Lee12,Barstow13g,Barstow16} or more recently \citet{Shulyak19}.
Bayesian methods such as optimal estimation (OE) heavily rely on a priori knowledge that is in general readily available for Earth's atmosphere as well as many Solar System bodies, but which is less well-known for exoplanets.
Grid based search methods have become a standard approach for exoplanet characterisation;
because of the need to perform millions to billions of radiative transfer forward model runs, sophisticated methods based on Monte Carlo Markow Chains and Nested Sampling are exploited to speed up the search
\citep[for recent reviews see e.g.][]{Madhusudhan18,BarstowHeng20,MacDonald23}.

\citet{Line13} have performed an intercomparison of three retrieval codes utilising OE and two Monte Carlo methods and concluded that for good measurements (high spectral resolution and little noise) the estimates agree quite well.
More recently, \citet{Barstow20r} compared three retrieval codes CHIMERA \citep{Line12}, NEMESIS \citep{Irwin08}, and Tau-REx \citep{Waldmann15t} and reported mostly consistent results
but emphasised the important role of radiative transfer modeling since the same inverse problem solver \citep[nested sampling MultiNest,][]{Feroz09} is used by all three codes; hence the testing of additional inverse solvers is desirable.
Likewise, the critical impact of the forward model was also emphasised by \citet{Barstow22} who presented an intercomparison of five codes (\citealp[ARCiS,][]{Min20}, NEMESIS, \citealp[Pyrat BAY,][]{Cubillos21}, Tau-REx, and \citealp[POSEIDON,][]{MacDonald17}).

Pressure, temperature, and molecular concentrations depend on space (and time), and a discretisation is mandatory for numerical analysis.
The latitudinal and longitudinal dependencies are largely ignored
(for a discussion of 2D or 3D effects on emission and transmission see e.g. \citealt{Feng16,Blecic17,Taylor20} and \citealt{Caldas19,MacDonald20}, respectively; see also \citealt{Pluriel23} for a recent review).
Whereas assuming molecular concentrations which are constant in altitude is likely acceptable as a first step, the assumption of isothermal temperature profiles \citep[e.g.][]{Barstow22} is problematic.
Layer-by-layer representations (standard for Earth remote sensing) have been criticised as troublesome due to the limited information content \citep{Line13,Parmentier14},
and parameterised representations of the vertical dependence of temperature are quite common \citep[e.g.][]{Madhusudhan09,Paris13r,Morley17gj}.
A novel function expansion approach along with a standard nonlinear least squares solver has been studied by \citet{Schreier20t}.
\citet{BarstowHeng20} recommended to ``conduct retrievals \dots\ with a variety of temperature structures and \dots\ to investigate alternative approaches.''

Here we examine the feasibility of temperature sounding by means of an iterative relaxation developed in the late sixties by \citet{Chahine68,Chahine70,Chahine72}
that is presented in several textbooks on atmospheric radiation \citep[e.g.][]{Goody89,Liou80,Hanel03,Zdunkowski07} but rarely used today.
According to ``Subsection 6.5.2 --- A physical approach to retrieval'' in \citet{Goody89} it is a ``simple idea easy to visualize and to extend to new circumstances'' and exploits the properties of the weighting functions
(the derivatives of the transmission w.r.t.\ altitude).
Extensions and/or refinements of this approach were presented by \citet{Smith70} and \citet{Twomey77a}.
These relaxation methods have been used for analysis of IR and microwave observations of Earth's atmosphere, and for temperature sounding of Venus \citep[e.g.][]{Taylor80}, Mars \citep{Lellouch91a, Lellouch91b, Haus00t}, and Jupiter \citep{Gautier77, Gautier79j}.

The organisation of the paper is as follows:
The following section describes our methodology (radiative transfer, Chahine relaxation, \dots) along with the code and data.
We demonstrate the feasibility to estimate the temperature for Earth-like exoplanets in Section \ref{sec:results}:
Computationally fast estimates exploiting a ``mapping'' of the wavenumber space to the altitude space as well as iterative refinements based on a comparison of observed and model spectra are presented.
We continue with a discussion in Section \ref{sec:disc} and give our conclusions in Section \ref{sec:concl}.


\begin{figure*}
 \centering\includegraphics[width=\textwidth]{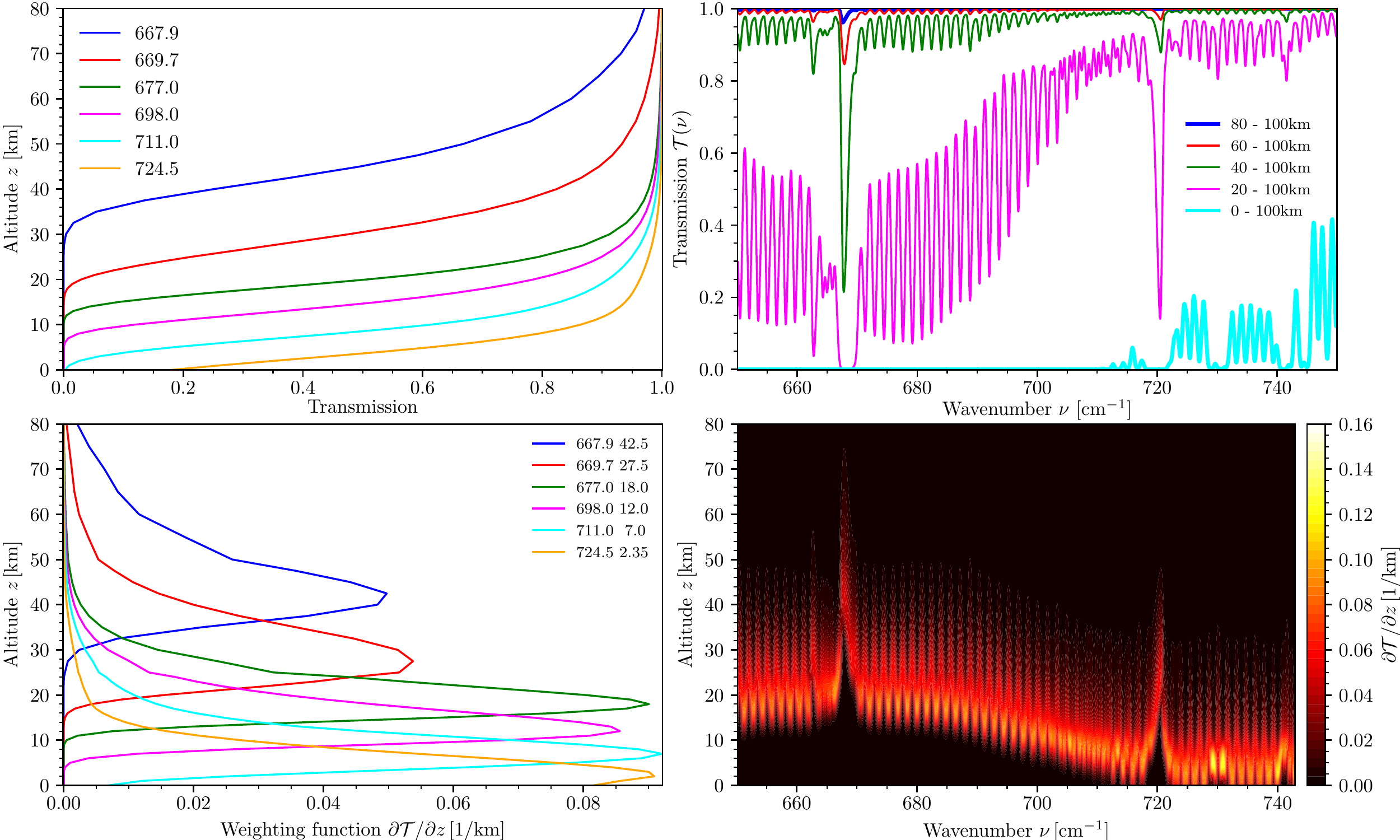}
 \caption{Transmission and weighting functions in the longwave TIR (\element{CO_2} $\nu_2$ band at $15\rm\,\mu m$, nadir view, Earth's midlatitude summer (MLS) atmosphere \citep{Anderson86},
          Gaussian response function of half width $\Gamma=0.25\rm\,cm^{-1}$ corresponding to a resolution $R \approx 2800$.).
          The upper panels show the transmission as a function of wavenumber for several atmospheric paths (right) and transmission vs.\ altitude (related to path length $s=z_\text{ToA}-z$) for selected wavenumbers (left).
          The lower panel shows individual weighting functions for selected wavenumbers (left) and a contour plot.
          Numbers in the left legends indicate the wavenumber [cm$^{-1}$] and the corresponding peak altitude [km] (bottom-left).}
 \label{wgtFct_tir2}
\end{figure*}

\section{Theory}
\label{sec:theory}

\subsection{Forward model --- infrared radiative transfer}
\label{ssec:irrt}

In a gaseous atmosphere with local thermodynamic equilibrium the upwelling intensity (radiance) $I$ at wavenumber $\nu$ is described by the Schwarzschild equation of radiative transfer \cite[]{Goody89,Hanel03}
\begin{align} \label{schwarzschild}
 I(\nu) ~&=~  \T(\nu,0) \; B(\nu,T_\text{surf})  ~+~ \int_0^{\tau(\nu,z)} B(\nu,T(\tau')) ~ \exp{\bigl(-\tau'(\nu)\bigr)} \, \D \tau' \\
\label{schwarzschildWgtFct}
        ~&=~  \T(\nu,0) \; B(\nu,T_\text{surf})  ~-~ \int_0^\infty B(\nu,T(z')) ~ {\upartial \T(\nu,z') \over \upartial z'} \, \D z'
\end{align}
where $B = 2 hc^2 \nu^3 \big/ \left[ \exp{(hc \nu / k_\text{B} T)} -1 \right]$ is Planck's function at temperature $T$ ($h$, $c$, $k_\text{B}$ are the Planck constant, speed of light, and Boltzmann constant, respectively).
For the surface contribution $I_\text{surf}(\nu) ~\equiv~ \T(\nu,0) \; B(\nu,T_\text{surf})$ we assume a Planck black body emission attenuated by the intermediate atmosphere 
and a temperature identical to the bottom of the atmosphere (BoA) temperature, i.e.  $T_\text{surf}=T_\text{BoA}=T{(z=0)}$.

The monochromatic transmission $\T$, closely related to the optical depth $\tau$, between observer and altitude $z$ is given by Beer's law
\begin{align} \label{beer}
 \T (\nu,z) ~&=~ \exp\bigl(-\tau(\nu,z)\bigr) \\ \notag
             &=~ \exp \left[ - \int_z^\infty \sum_m k_m\bigl(\nu,p(z'),T(z')\bigr) \: n_m(z') \; dz' \right] ~,
\end{align}
with $n_m$ the density of molecule $m$, and $k_m$ the pressure and temperature dependent absorption cross section obtained by summing over the contributions from many lines.
For simplicity a vertical path is assumed; for a slant path with angle $\theta$ in a plane-parallel atmosphere replace $z' \longrightarrow z'/\cos(\theta)$. 
The finite spectral resolution of the instrument is taken into account by convolution of the monochromatic intensity \eqref{schwarzschild} (or transmission \eqref{beer}) with a spectral response function (SRF, e.g. Gaussian).

The upper atmosphere has a low abundance of absorbers and is therefore almost transparent (i.e., transmission close to one);
with an increasingly longer atmospheric path (decreasing $z$) attenuation becomes stronger especially at wavenumbers with strong absorption in the band or line center (\qufig{wgtFct_tir2} top-right).
Here path length $s$ refers to the distance to the observer, essentially at ``infinity'', in practice at top-of-atmosphere (ToA), and is linked to altitude via $s=z_\text{ToA}-z$, cf.\ \EqRef{beer}.
Viewed as a function of altitude the transmission decays rapidly to zero for these wavenumbers (\qufig{wgtFct_tir2} top-left), i.e.\ photons from the lower atmosphere cannot penetrate to space and the ToA radiation arises mainly from the upper atmospheric layers.
The so-called weighting function,%
\footnote{A note on terminology:
different terms are used for the derivative $\upartial \T / \upartial z$: weighting functions \citet{Liou80,Hanel03,Zdunkowski07}, kernel functions \citet{Goody89}, or contribution functions.
Moreover, ``Jacobian'' and ``weighting function'' are often used interchangeably; however, for temperature retrievals using nonlinear least squares the Jacobian is the partial derivative of the radiance w.r.t.\ the state vector $\vec x$,
$\upartial I / \upartial x$, where $\vec x$ is a discrete representation of the temperature profile.}
the partial derivative $K(\nu,z) \equiv \upartial \T(\nu,z) / \upartial z$ in \eqref{schwarzschildWgtFct}, quantifies the dominant contribution (or weight) of an altitude layer to the outgoing radiation (\qufig{wgtFct_tir2} bottom panels, see also \citet[][Fig.\ 7.6]{Liou80}, \citet[][Fig.\ 6.17]{Goody89}, and \citet[][Fig.\ 8.2.1]{Hanel03}).
Note that the weighting function is similar, but not identical to the temperature Jacobian $\upartial I(\nu) / \upartial T(z)$ that measures the radiation's sensitivity to changes of temperature \citep[see][Fig.\ 10]{Schreier20t}.
Both the weighting functions and the Jacobian clearly demonstrate that the radiation carries little information of the lowermost and upper atmospheric layers.

\begin{figure*}
 \centering\includegraphics[width=\textwidth]{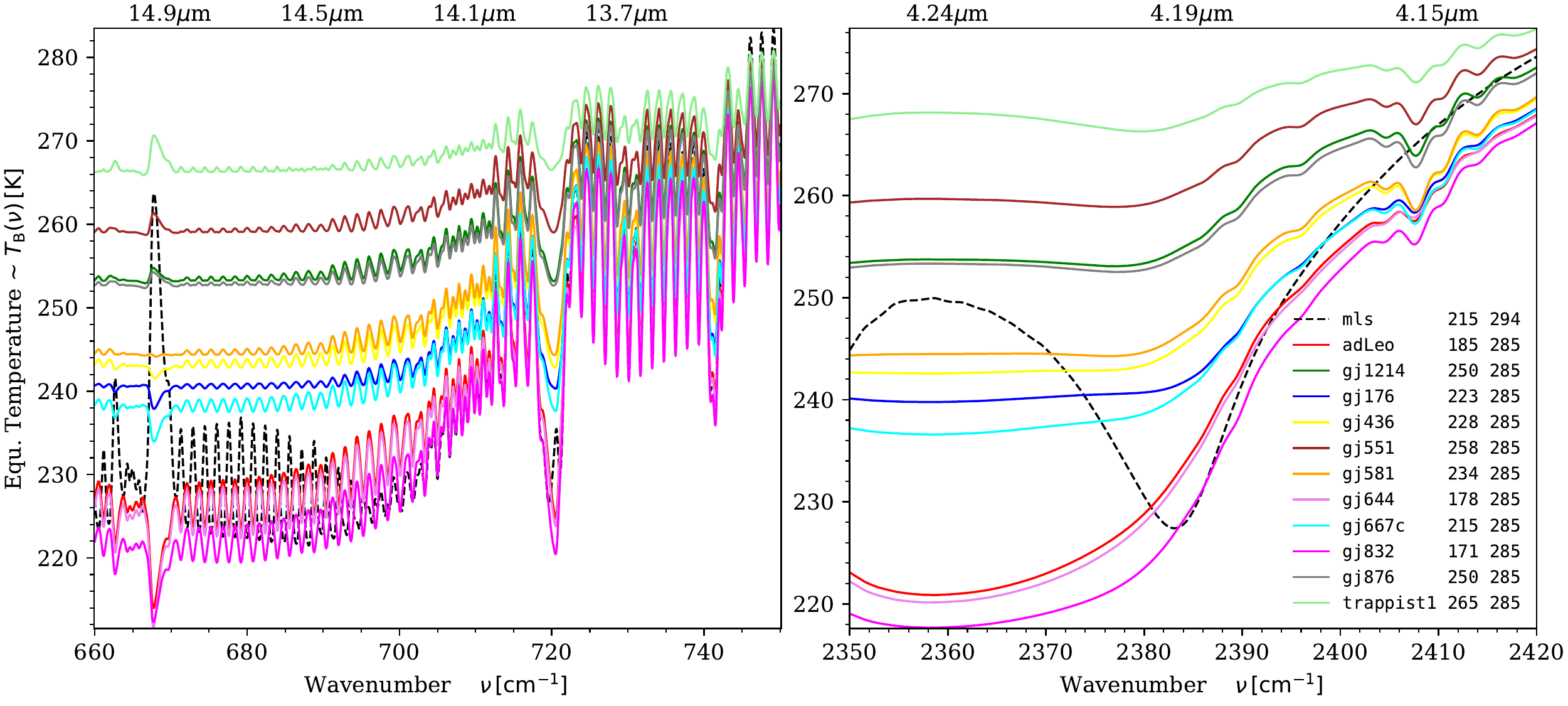}
 \caption{``Observed'' equivalent brightness temperatures $T_\text{B}$ \eqref{equTempGuess} of hypothetical Earth-like planets orbiting the M-dwarfs indicated in the legend. Earth's MLS atmosphere is shown for comparison (see subsection \ref{ssec:data}).
          Gaussian response function of HWHM $\Gamma=0.25\rm\,cm^{-1}$ (longwave, left) and $\Gamma=1.0\rm\,cm^{-1}$ (shortwave, right).
          For clarity no noise has been added. The TIR-LW and TIR-SW spectra comprise 1441 and 281 pixels, respectively.
          The numbers in the legend list the minimum and maximum atmospheric temperature [K].
          Wavelengths are given at the top (numerically $\lambda[\mue]=10^4/\nu[\cm]$).
          }
 \label{observedBT}
\end{figure*}

\subsection{Inversion --- Chahine relaxation}
\label{ssec:inv}

According to \citet{Rodgers76} ``the intensity to be measured is \dots\ a weighted mean of the Planck function profile with the weighting function''.
The bell-shape of the weighting function (\qufig{wgtFct_tir2} bottom-left) can be exploited for analysis of TIR spectra.
Assuming a delta-function-like weighting function the Schwarzschild equation \eqref{schwarzschild} reduces \citep{Hanel03} to
\begin{equation} \label{schwarzschildApprox}
    I(\tilde\nu) ~\approx~ I_\text{surf}(\tilde\nu) ~+~ B\bigl(\tilde\nu, T(\tilde z_\nu)\bigr)
\end{equation}
with $\tilde z_\nu$ the altitude where $K(\nu,z)$ has a maximum for a given $\nu$.
A first approximation of the atmospheric temperature can thus be inferred from the observed Equivalent Brightness Temperature (EBT), i.e.
\begin{equation} \begin{aligned} \label{equTempGuess}  
    T(\tilde z_\nu) ~\approx~ T_\text{B}(\tilde\nu) ~&\equiv~ B^{-1}\Bigl( I_\text{obs}(\tilde\nu) \,-\, I_\text{surf}(\tilde\nu) \Bigr) \\
     &=~ {hc\tilde\nu/k_\text{B} \over \log\Bigl( {2 h c^2 \tilde\nu^3 / \bigl(I_\text{obs}(\tilde\nu) \,-\, I_\text{surf}(\tilde\nu)\bigr)} \Bigr)}
\end{aligned} \end{equation}
This estimate can be iteratively improved using a relaxation scheme originally proposed by \citet{Chahine68,Chahine70}
\begin{equation} \label{chahineRelax}  
    T_{i+1}(\tilde z_\nu) ~\approx~ B^{-1}\left( {I_\text{obs}(\tilde\nu)\,-\,I_\text{surf}(\tilde\nu) \over I_\text{mod}(\tilde\nu, T_i)\,-\,I_\text{surf}(\tilde\nu)} ~ B\bigl(\tilde\nu,T_{i}(\tilde z_\nu) \bigr) \right)
\end{equation}
where $T_i$ denotes the temperature for iteration $i$ and $I_\text{mod}(\tilde\nu, T_i)$ the corresponding modelled radiance according to \eqref{schwarzschild}.

\begin{figure}
 \centering\includegraphics[width=\linewidth]{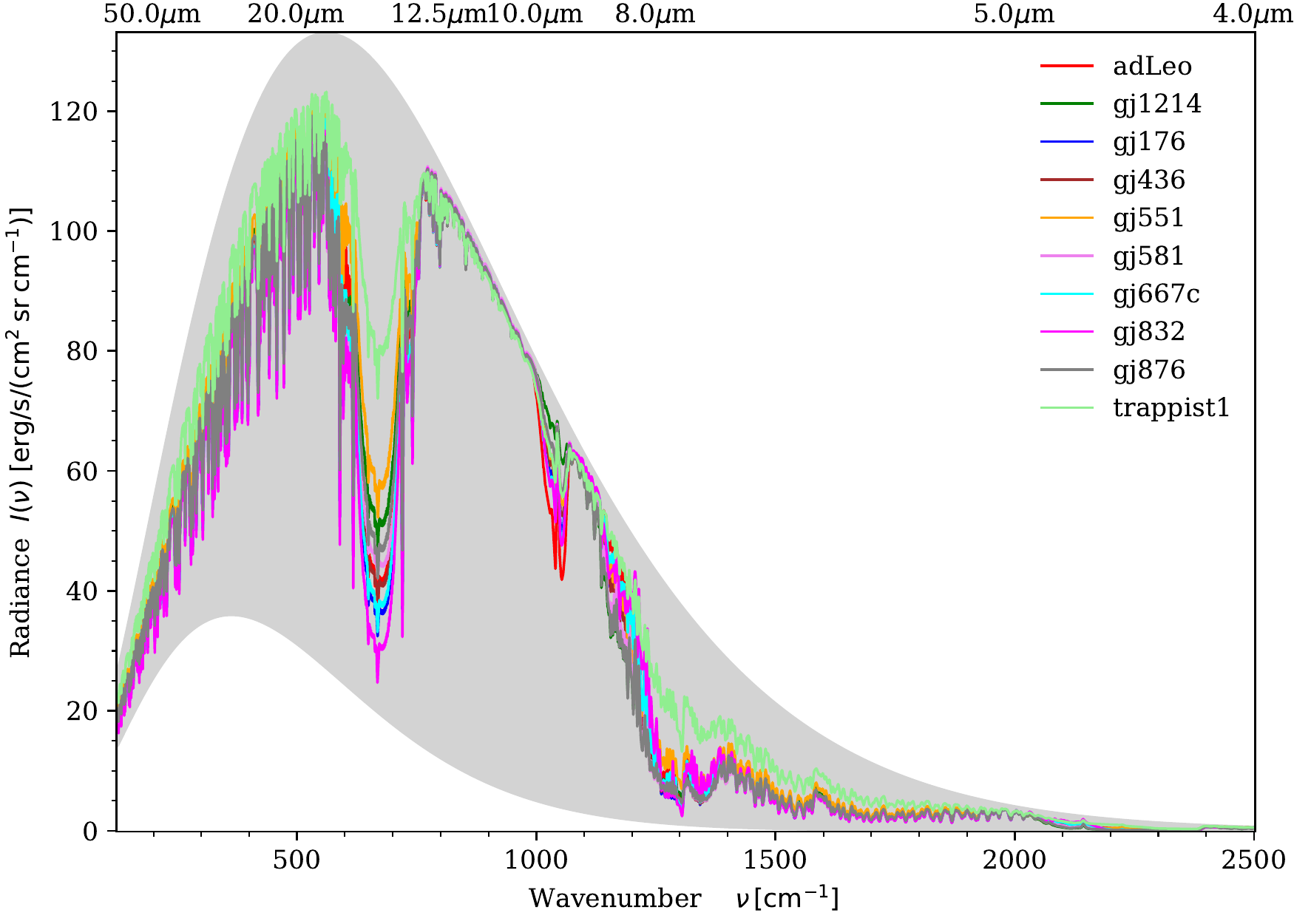}
 \caption{ToA intensities for M-Earths. The gray-shaded area indicates the Planck function $B(\nu,T)$ for the minumum and maximum atmospheric temperatures $T_\text{min}\approx 171\rm\,K$ for GJ\:832 and $T_\text{max}\approx 286\rm\,K$ by ``construction''. (Computed with GARLIC using atmospheric data from 1D-TERRA \citep{Wunderlich20}; Gaussian response function with resolution $R=1000$.)}
 \label{radiancesToA}
\end{figure}

\subsection{Implementation}
\label{ssec:code}

For our forward model we use Python for Computational ATmospheric Spectroscopy \citep[Py4CAtS, ][available at \url{https://atmos.eoc.dlr.de/tools/py4cats/}]{Schreier19p}, a Python re-implementation of the Generic Atmospheric Radiation Line-by-line Infrared Code \citep{Schreier14}.
GARLIC has been thoroughly verified by intercomparison with other codes \citep[e.g.][]{Schreier18agk} and validated by comparison to effective height spectra \citep{Schreier18ace} generated from Earth observations of the Atmospheric Chemistry Experiment --- Fourier transform spectrometer \citep[ACE-FTS,][]{Bernath17}.

Py4CAtS and GARLIC compute molecular absorption cross sections $k_m$ assuming a (default) Voigt line shape \citep{Schreier18h}, where the wavenumber grid point spacing is 
adjusted automatically for each molecule, pressure and temperature to a fraction of the typical line width.
Next, cross sections scaled by molecular number densities are summed up to absorption coefficients; then standard quadrature schemes are used to compute optical depths and radiances.
Both observed and modeled spectra are convolved with a Gaussian spectral response function of constant width (with a default sampling of 4 points per half width at half maximum (HWHM)).
Py4CAtS makes heavy use of Numpy \citep{vanderWalt11,Harris20} (and occasionally SciPy \citep{Virtanen20} and MatPlotLib \citep{Hunter07}).

The synthetic measurement spectrum is generated by adding generic Gaussian noise (generated by the Numeric Python \texttt{numpy.random.randn} function) independent of wavenumber,

The equivalent brightness temperature spectrum is obtained by ``inversion'' \eqref{equTempGuess} of Planck's function and depicted in \qufig{observedBT} for noise-free simulated observations (see next subsection).

\begin{figure*}
 \centering\includegraphics[width=\textwidth]{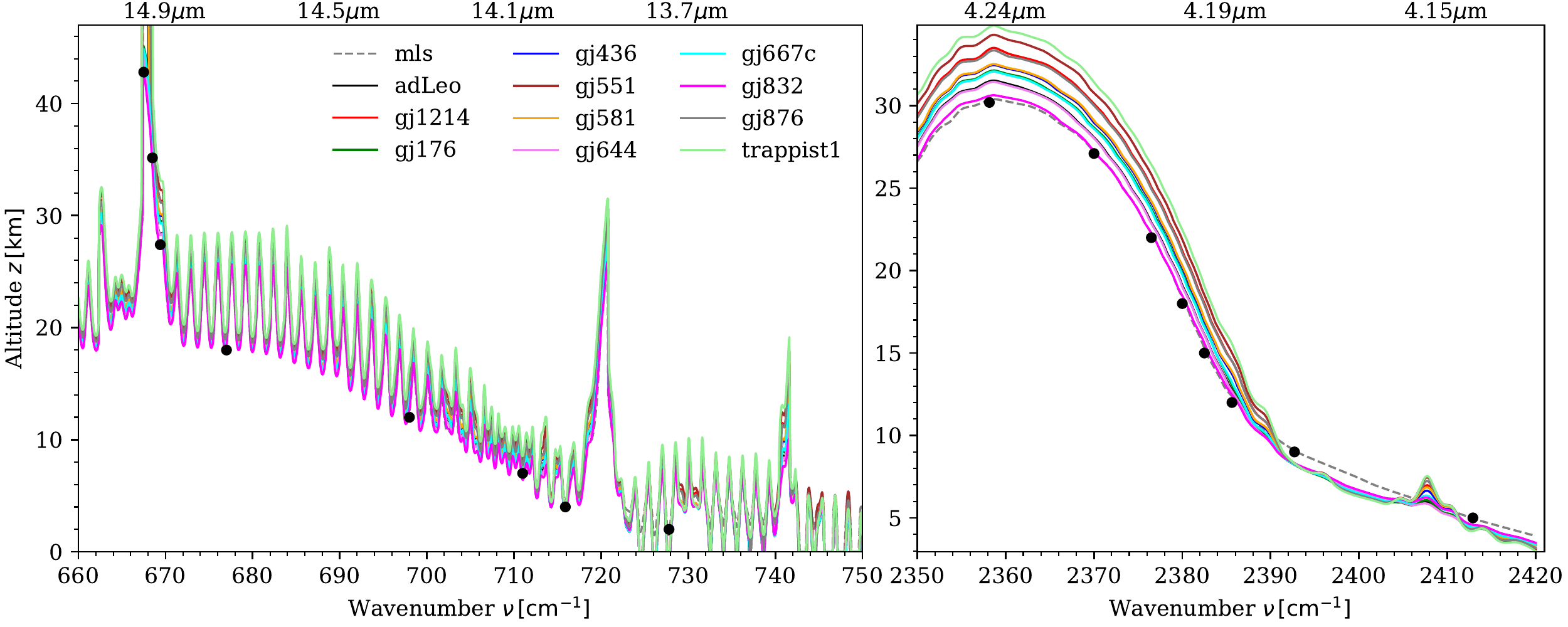}
 \caption{Weighting function maxima in the longwave TIR (\element{CO_2} $\nu_2$ band at $15\rm\,\mu m$ with Gaussian response function of half width $\Gamma=0.25\rm\,cm^{-1}$, left) and
                                          shortwave TIR (\element{CO_2} $\nu_3$ band at $4.3\rm\,\mu m$ with $\Gamma=1.0\rm\,cm^{-1}$, right).
          The length of these spectra (i.e.\ number of data points) is identical to those of \qufig{observedBT}.
          The black dots indicate manually selected $\nu \leftrightarrow z$ pairs (as also shown in \qufig{wgtFct_tir2} lower left).}
 \label{wgtFct_maxima}
\end{figure*}

\subsection{Data}
\label{ssec:data}

Atmospheric data for Earth are taken from the ``AFGL Atmospheric Constituent Profiles'' atmospheres \citep{Anderson86} providing pressure and temperature vs.\ altitude along with concentration profiles for 28 gases including water vapor, carbon dioxide (with 360\,ppm volume mixing ratio (VMR)), ozone, methane.

Atmospheric data for the assumed Earth-like planets around M-dwarfs (henceforth called ``M-Earths'') are taken from \citet{Wunderlich19}.
These temperatures and concentrations are inferred from a 1D photochemistry model \citep{Gebauer18e} coupled to a climate model and are defined on 64 levels with a ToA pressure of about $0.08\rm\,mb$.
The former work assumed hypothetical planets with Earth's properties orbiting different M-dwarf stars placed at the location where the planet receives modern Earth's instellation.
For all M-Earths the surface temperature is approximately 288\,K by appropriate selection of the orbital radius,
and the \element{CO_2} VMR is about 355\,ppm in the lower atmosphere.

Data for Trappist-1e from \citet{Wunderlich20} are derived from the Berlin 1D steady-state, cloud-free, radiative-convective photochemical model 1D-TERRA.
This dataset comprises dry\&dead, wet\&dead, and wet\&live scenarios for \element{CO_2} surface partial pressures of $10^{-3}$, $10^{-2}$, $10^{-1}$, and $10^{0}\rm\,bar$
(corresponding to VMRs of approximately $10^{-3}$, $10^{-2}$, $10^{-1}$, and $0.5 \cdot 10^{0}$ (see Table 11 in \citet{Wunderlich20} and is indicated by the exponent in subsection \ref{ssec:Trappist1}).
For comparison, the Trappist-1 planet of the M-Earth dataset \citep{Wunderlich19} has a VMR of 355\,ppm in the lower atmosphere.
Note that this planet is a purely hypothetical ``Earth'' orbiting Trappist-1, whereas the Trappist-1e scenarios are based on orbital and stellar data \citep[see Section 3.2 and Table 9 in][]{Wunderlich20}.

The Chahine approach delivers temperatures only at a small set of altitudes, but data on a moderately dense grid from BoA to ToA are required for the radiative transfer modeling and we will use function expansion for inter/extrapolation (see subsection \ref{ssec:chahine}).
For the generation of the singular vectors to be used as base vectors for this expansion we use the set of 42 Earth atmospheric profiles collected by \citet{Garand01} augmented with the eleven M-Earth temperatures regridded to a uniform altitude grid with 2\,km steps \citep[][subsection 3.2]{Schreier20t}.
Note that the first six Garand atmospheres correspond to the AFGL data.

Molecular line parameters are taken from the \textsc{Hitran} database;
instead of the most recent data \citep{Gordon22} (clearly mandatory for analysis of real observations) we use data from the initial 1986 release \citep{Rothman87} to speed-up the computations.
(See the further discussion in subsection \ref{ssec:discHit}.)
Only the main IR absorbers are considered, i.e.  \element{CO_2} and the interfering species \element{H_2O}, \element{CH_4}, and \element{O_3}.

See also \citet{Schreier20t} for more details on atmospheric and molecular data and a discussion of some of our approximations and assumptions.


\section{Results}
\label{sec:results}

\subsection{First preliminary constraints}
\label{ssec:prelim}

Inspection of the Schwarzschild equation \eqref{schwarzschild} can be used for a first estimate of the range of atmospheric temperatures.
Replacing the height-dependent temperature $T(z)$ in the Planck function by the minimum atmospheric value $T_\text{min}$, the equation simplifies to
$I(\nu) = B_{BoA} \E^{-\tau} + \int B \E^{-\tau'} \D \tau' \ge B_{min} \left(\E^{-\tau} + (1-\E^{-\tau'}) \right) = B_\text{min}$,
hence the ToA radiance is greater (or equal) to the minimum Planck function.
The upper limit can be derived in a similar manner, hence $B_\text{min} \le I(\nu) \le B_\text{max}$.

The intensities shown in \qufig{radiancesToA} confirm these constraints, i.e.\ the lower and upper ``Planck envelope'' can be used as a preliminary estimate of the temperature range.
The minimum and maximum atmospheric temperatures can then be readily estimated from the corresponding EBT minima and maxima.

The EBT spectra in \qufig{observedBT} are, except for resolution, essentially a zoom-in on the intensity spectra of \qufig{radiancesToA} transformed via \EqRef{equTempGuess}.
In addition to the strong absorption at $15\rm\,\mu m$ due to \element{CO_2} the ozone fundamental band at $1042 \cm$ ($9.6\mue$) is clearly visible in \qufig{radiancesToA}.

\subsection{Mapping Wavenumbers to Altitudes}
\label{ssec:vzMap}

Before estimating atmospheric temperatures from IR spectra according to the recipe of \EqRef{equTempGuess} several issues have to be addressed.
First a set of appropriate wavenumber-altitude pairs has to be identified.
Suitable wavenumbers are conveniently searched for in absorption band(s) of a molecule with well-known concentration and ideally little variability, e.g.\ the shortwave or longwave TIR bands of \element{CO_2}.
The corresponding altitudes are given by the location of the weighting function maxima, hence depend on the properties of the transmission $\T$ and as a consequence depend on atmospheric temperature, pressure, and composition.
Obviously these data are unknown for exoplanets, but fortunately weighting function of Earth-like exoplanets (with \element{N_2}-\element{O_2} dominated atmospheres) are rather similar and closely resemble weighting functions of typical Earth climates, see \qufig{wgtFct_maxima}.
Accordingly we will use the $\nu \leftrightarrow z$ mapping of Earth's atmosphere, e.g.\ for midlatitude and subarctic summer or winter (MLS, MLW, SAS, SAW) in the $\nu_2$ longwave (LW, wavelength $\lambda \approx 15\rm\,\mu m$) and $\nu_3$ shortwave (SW, $\lambda \approx 4.3\rm\,\mu m$) bands of \element{CO_2}.
Note that these spectra also depend on resolution (compare Figures 1 and 2 of \citet{Schreier20t}), hence a lower resolution leads to smoother spectra and reduces the sensitivity to upper atmospheric layers
(See the discussion in subsection \ref{ssec:resolution}).

\begin{figure}
 \centering\includegraphics[width=\linewidth]{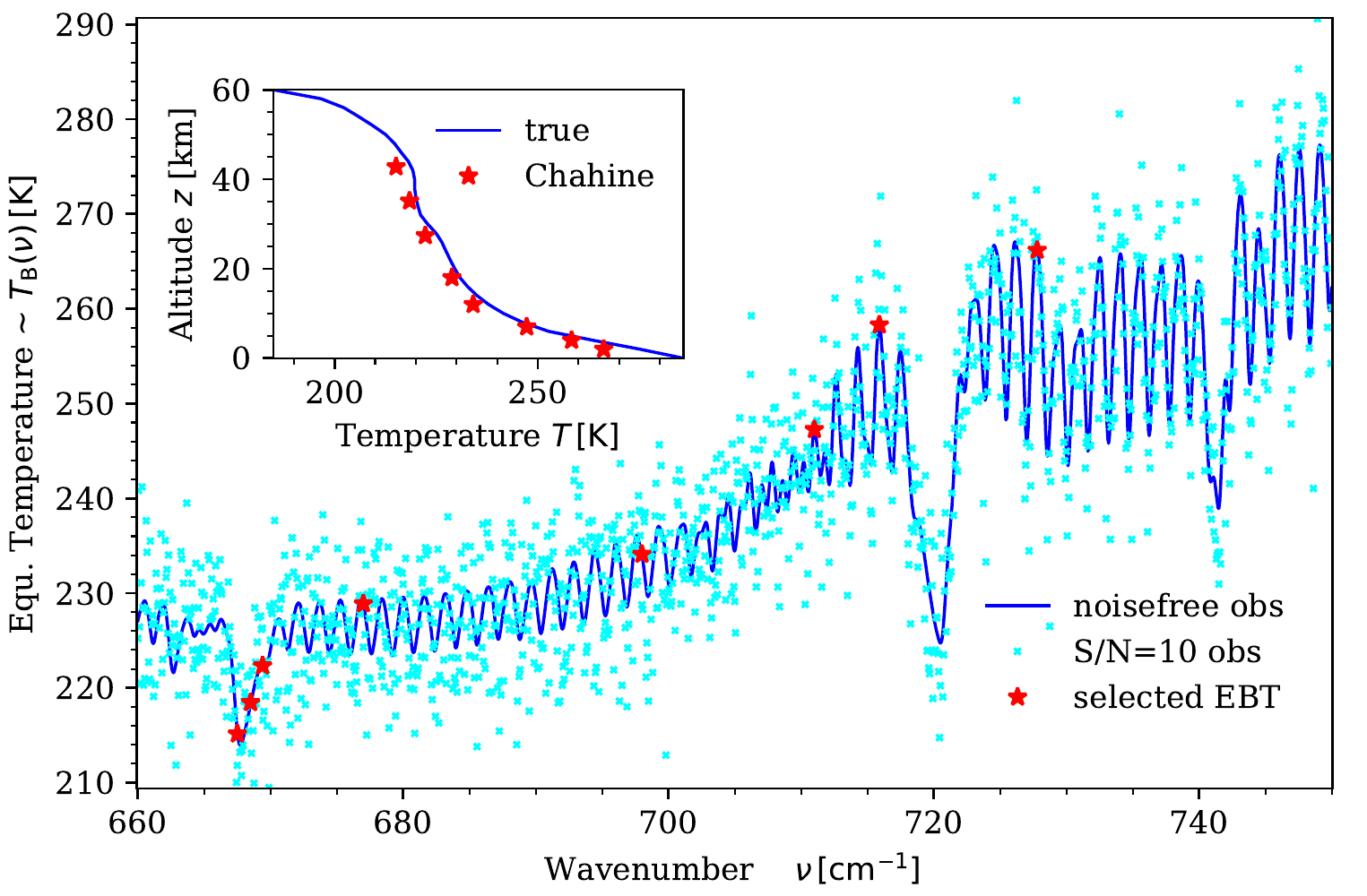}
 \caption{Estimate of AD\:Leo atmospheric temperature using a $\nu \leftrightarrow z$ mapping with 8 pairs for MLS (see \qufig{wgtFct_maxima}).
          The main plot shows the ideal noise-free ``observed'' EBT (blue), the noise contaminated EBT spectrum ($S/N=10$, cyan), and selected EBT values taken from the noise-free spectrum;
          the inset compares the estimated temperatures with the true profile.
          (TIR-LW $660 \text{--} 750\rm\,cm^{-1}$; Gaussian with $\Gamma=0.25\rm\,cm^{-1}$.) }
 \label{firstGuess8adLeo}
\end{figure}

\subsection{First Guesses --- Selected Data in TIR-LW}
\label{ssec:firstGuess8}

Having identified the translation from wavenumber to altitude space (henceforth called ``mapping'') it appears to be straightforward to infer the temperatures from the observed spectrum using \EqRef{equTempGuess}.
However, both real spectra as well as our theoretical spectra simulating planned instrumental measurements (see \qufig{observedBT}) are contaminated by noise, and exploiting single data pairs $(\tilde\nu, T_\text{B})$ is likely to lead to a ``noisy'' temperature profile.
This is confirmed by \qufig{firstGuess8adLeo} where the temperature retrieval for a hypothetical Earth-like planet orbiting AD\:Leo \citep{Wunderlich19} is illustrated:
the inset shows the eight temperature values taken from the observed noise-free observation (intensity spectrum $I(\nu)$ converted to EBT $T_\text{B}(\nu)$ according to \eqref{equTempGuess}).
However, estimating these temperatures from the noisy EBT spectrum would clearly lead to a zigzag temperature profile.

\begin{figure}
 \centering\includegraphics[width=\linewidth]{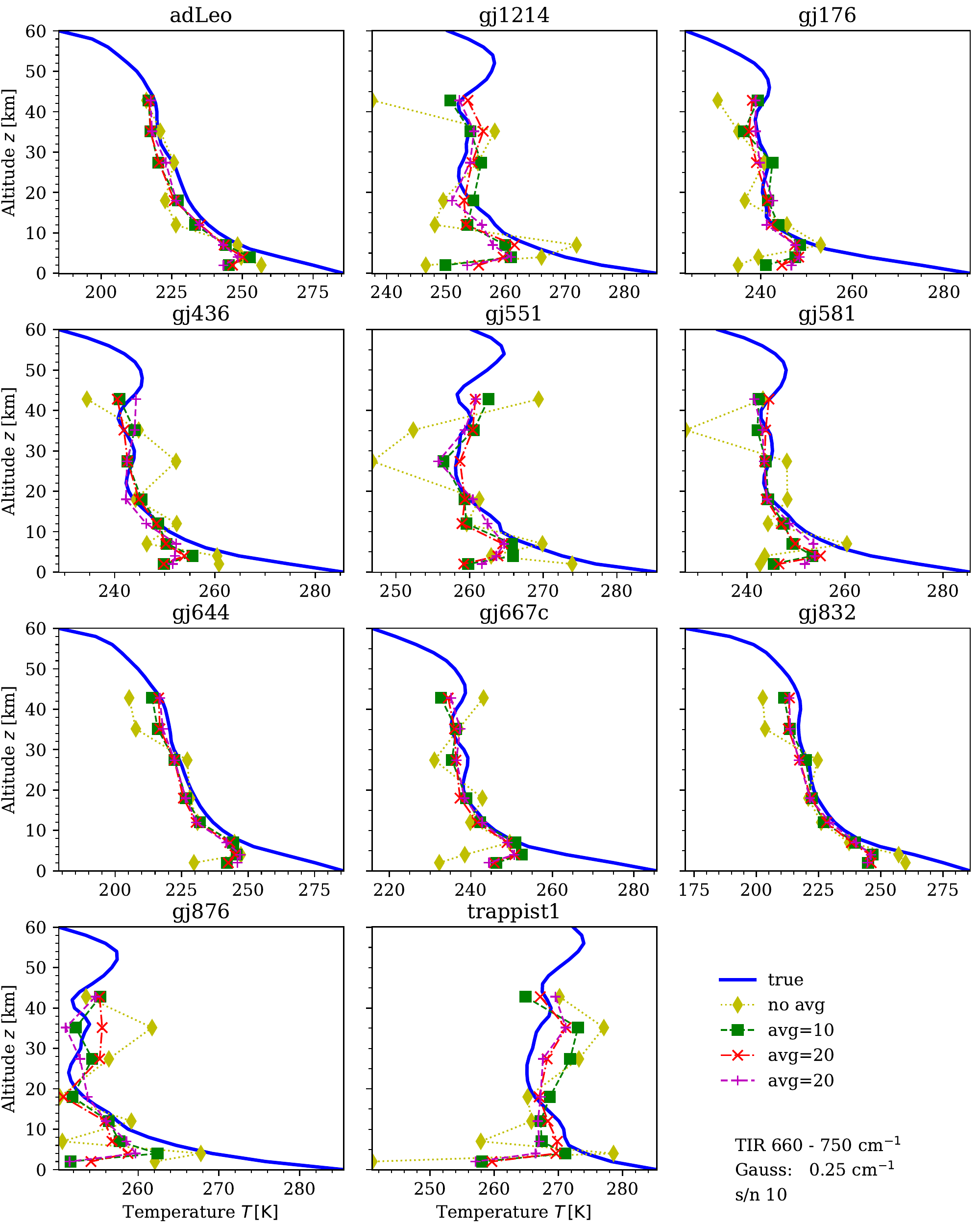}
 \caption{First estimate of M-Earths' atmospheric temperatures using a $\nu \leftrightarrow z$ mapping with 8 pairs for MLS (see \qufig{wgtFct_maxima}).
          The red and magenta curves labeled ``avg=20'' are identical except for the different noise vector.
          (Spectra as in \qufig{firstGuess8adLeo})}
 \label{firstGuess8}
\end{figure}

To compensate for the noise the average of some neighboring pixels from the observed EBT spectrum can be used instead.
Figure \ref{firstGuess8} depicts temperature estimates for the eleven M-Earth atmospheres (taken from \citet{Wunderlich19} and already used in \citet{Schreier20t})
and clearly demonstrates that averaging EBTs from larger windows (e.g. 10 or 20 pixels) leads to the inferred temperature profile becoming smoother.
Averaging 20 pixels leads to a significantly reduced zigzag, however, profiles estimated from two different measurements (model spectrum contaminated by two randomly generated noise vectors) are still distinct.
For the M-Earths these first guess temperatures are encouraging, for some cases almost ``perfect'' (e.g.\ AD\:Leo, GJ\:644, and GJ\:832 even without averaging), but in a few other cases (e.g.\ GJ\:551, GJ\:876, and Trappist-1) deviations to the true profile are clearly visible.
Moreover, temperatures of Trappist-1 in the lowermost altitudes are significantly underestimated.

\begin{figure}
 \centering\includegraphics[width=\linewidth]{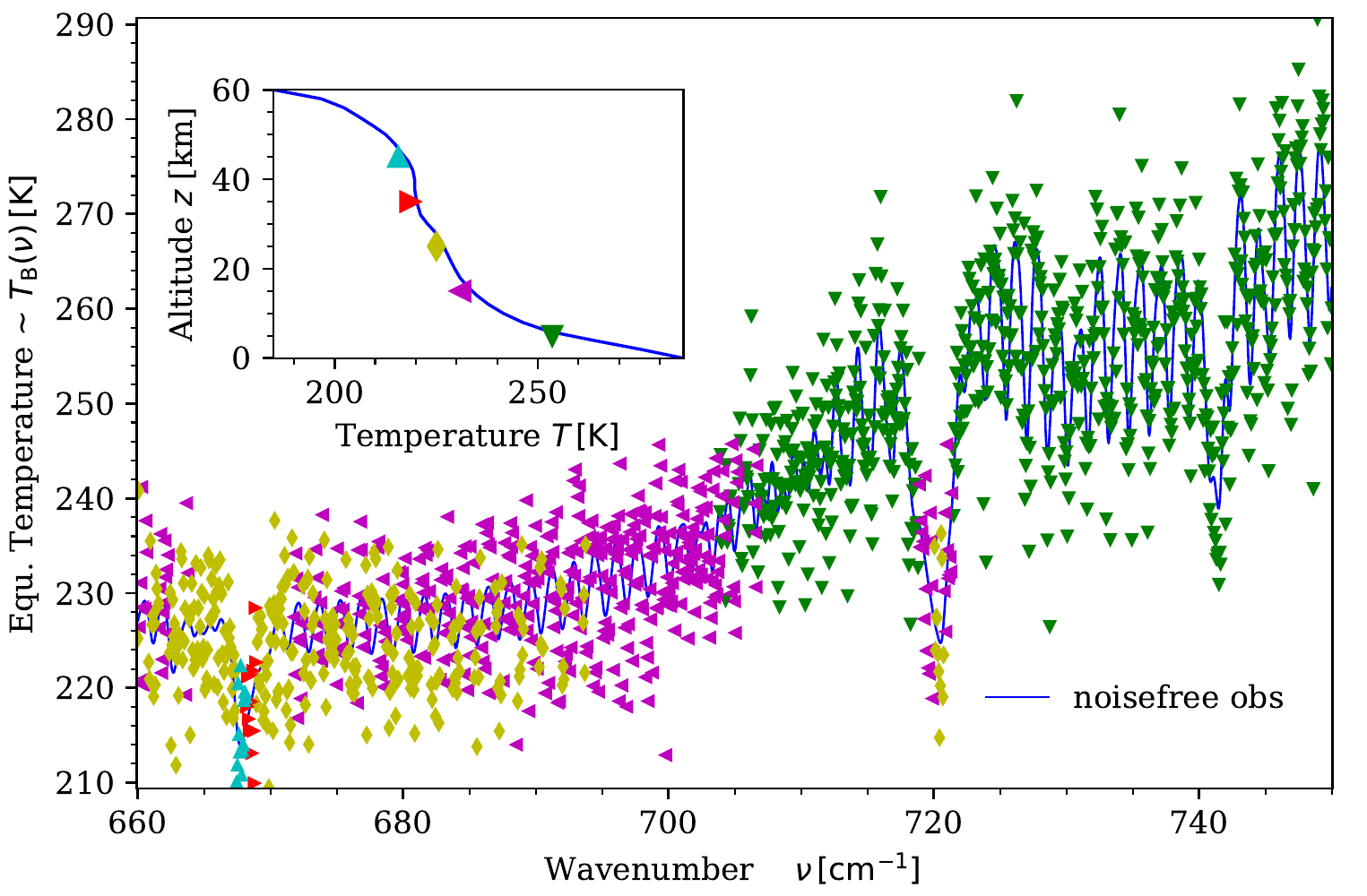}
 \caption{Estimate of AD\:Leo atmospheric temperatures using all $\nu \leftrightarrow z$ pairs for MLS.
          The colors and symbols for the $S/N=10$ observation indicate the altitude range where radiation is mainly coming from according to the weighting functions.
          (For example green and cyan triangles for the lowest and highest altitude. Inset similar to \qufig{firstGuess8adLeo}, spectra as above.)}
 \label{secondGuess_adLeo}
\end{figure}

The profiles of GJ\:551, GJ\:876, and Trappist-1 show oscillations of up to a few Kelvin.
The cold trap of Trappist-1 is not very strong, i.e.\ the minimum temperature is only 20\,K cooler than the maximum temperature (288\,K at BoA).
The tropopause temperatures of GJ\:551 and GJ\:876 are slightly cooler (258\,K, and 251\,K, respectively).
Moreover, the local temperature maximum of Trappist-1 in the mid atmosphere at about 38\,km is only modest and not reproduced by the EBT estimate.
Overall however, the results are rather encouraging.

\subsection{First Guesses --- Exploiting the entire TIR-LW}
\label{ssec:firstGuess}

Despite the promising results there are some caveats.
Exploiting just a few data points implies that the majority of data remain unused.
Moreover, the hand-picked selection of $\nu \leftrightarrow z$ mapping pairs is somewhat arbitrary, and the estimated atmospheric temperatures are therefore likely to change with different mappings.
Consider for example the mapping pairs $(677\rm\,cm^{-1},~18\,km), ~(698\rm\,cm^{-1},12\,km),  ~(711\rm\,cm^{-1},~7\,km)$ (cf.\ \qufig{wgtFct_maxima}):
One issue is whether the mapping wavenumber should be chosen to lie in the valley or on the peak of the absorption line.
Choosing a point somewhere in-between is less sensitive to resolution and less likely sensitive to noise.
Furthermore, the number of pixels which are used for averaging to account for the noise is limited, in order to avoid overlap of spectral regions.
This is especially the case near the band center.

Figure \ref{wgtFct_maxima} indicates that (except for the bottom and top altitudes) several wavenumbers are sensitive to a particular altitude.
For example, the atmospheric layer around 32\,km influences the radiance at several wavenumbers in the $660 \,\text{--}\, 670 \rm\,cm^{-1}$ interval and near $720 \rm\, cm^{-1}$ (for some planets only).
In order to address this issue, we therefore use the mean EBT of all pixels contributing to a particular altitude according to the weighting function peak height spectrum (\qufig{wgtFct_maxima}).
Because these peak altitudes rarely coincide exactly to a given grid point we accept all pixels within a given \emph{tolerance interval $\delta z$},
i.e.\ to estimate the atmospheric temperature at an altitude $z$ we take the average of all EBTs at wavenumbers with a weighting function peak height in the interval $[z-\delta z, z+\delta z]$.
This concept is illustrated in \qufig{secondGuess_adLeo} for a 5\,km tolerance:
Altitudes beyond 40\,km are seen only in a narrow interval around $667\cm$, and the average of all EBT's in this interval is interpreted as atmospheric temperature in the $40 \text{--} 50\rm\,km$ altitude range.
Likewise, temperatures for altitudes below 10\,km are estimated from EBT's at wavenumbers beyond $700\cm$ (except for the peak at $720 \rm\, cm^{-1}$). 

\begin{figure}
 \centering\includegraphics[width=\linewidth]{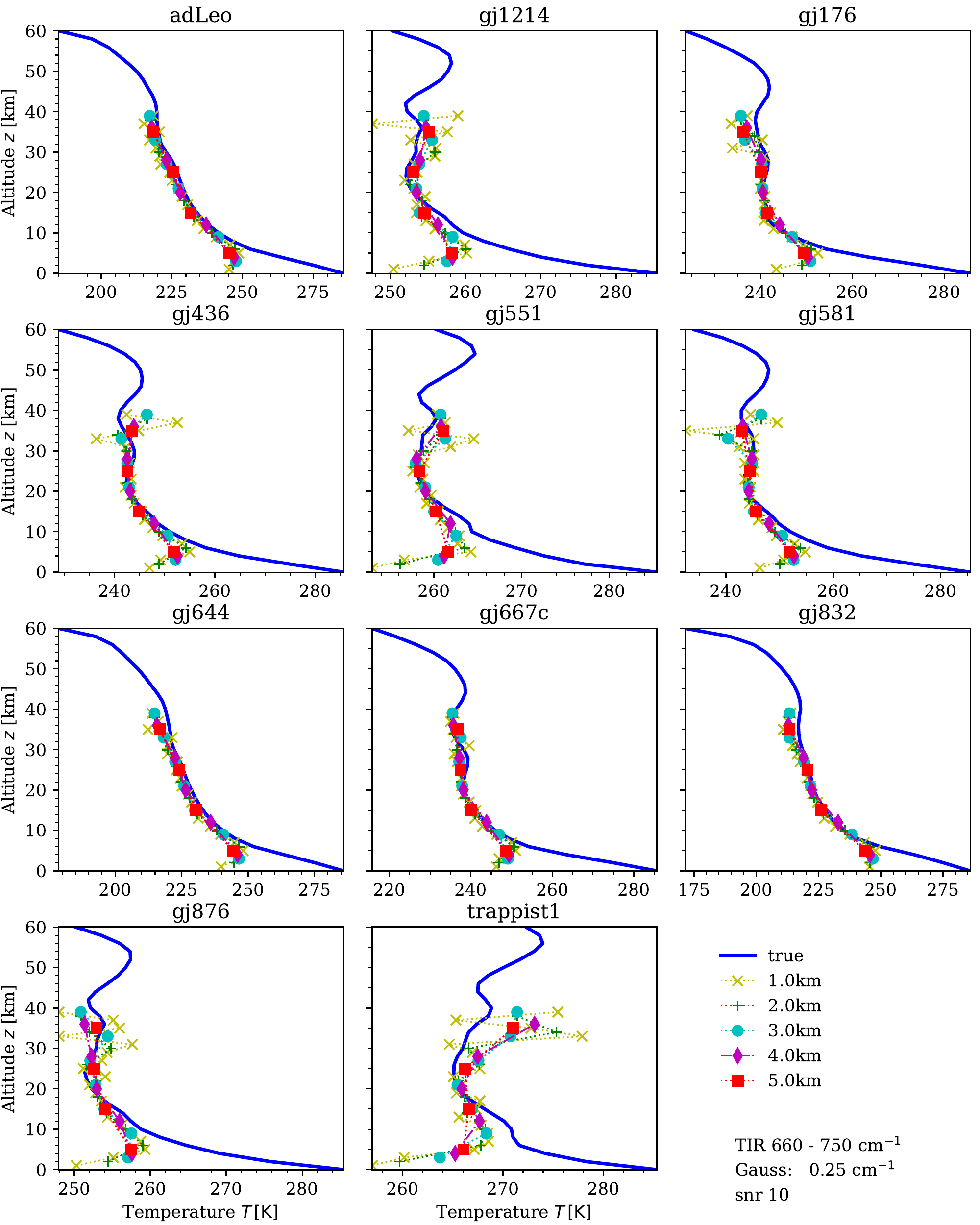}
 \caption{Estimate of M-Earths' atmospheric temperatures using all MLS $\nu \leftrightarrow z$ pairs.
          For the 1.0\,km and 5.0\,km tolerance mappings 18 and 4 altitudes have been found with $5 \le z \le 39\rm\,km$ and $5 \le z \le 35\rm\,km$, respectively.
          (Spectra as above.)}
 \label{secondGuessAll_TIR2}
\end{figure}

Figure \ref{secondGuessAll_TIR2} compares results for all M-Earths and for various $\delta z$-intervals.
In accordance with \qufig{wgtFct_maxima} no mappings are found for the upper atmosphere, and the highest altitude point estimated depends on the magnitude of the tolerance $\delta z$.
In the ``middle atmosphere'' the estimated temperatures are roughly equivalent.
In some cases (e.g.\ GJ\:664 or GJ\:832) where temperatures were slightly underestimated using 8 $(\nu,z)$ pairs only, the deviation is reduced or eliminated.
For the generalised Chahine case the retrieved temperatures are often smoother compared to \qufig{firstGuess8}.
Moreover, the extended retrievals are less sensitive to noise:
temperatures estimated with the 5\,km tolerance from two observations with different noise vectors are largely identical except for the highest values (at 35\,km) of GJ\:551 and Trappist-1
(compare the two ``avg=20'' estimates in \qufig{firstGuess8}).
The warm M-Earths (GJ\:551, GJ\:876, and Trappist-1) remain problematic with clear oscillations above about 30\,km and considerably underestimate near BoA especially for the 1 and 2\,km tolerances,
but otherwise the retrieved and true temperature profiles are in good agreement with deviations to $T_\text{true}$ less than ten Kelvin.

The oscillations of the profiles estimated with the 1 or 2\,km tolerances can be interpreted as follows:
A closer look to \qufig{wgtFct_tir2} shows that the bell-shaped weighting functions have a finite width of several kilometers in the lower atmosphere and almost 10 kilometers in mid to the upper atmosphere.
Hence retrieving temperatures with one kilometer resolution is questionable.

\subsection{Iterative Refinements --- TIR-LW} 
\label{ssec:chahine}

The temperature estimates presented in the previous subsection are not always satisfactory, but they can be used as initial guesses for the Chahine relaxation scheme \eqref{chahineRelax} or other iterative solvers like nonlinear least squares \citep[e.g.][]{Schreier20t}.
Clearly the main advantage is the computational speed, i.e.\ the whole ``retrieval'' is simply the inversion \eqref{equTempGuess} of the Planck function and therefore can be performed within fractions of a second. 
However, because the peak heights of the weighting functions do not cover the entire altitude range, this initial guess temperature values cannot readily be used as an input for radiative transfer modeling.
Moreover, with the $4$ or $5\rm\,km$ tolerances only few temperature values are estimated in the mid atmosphere.

Extrapolation appears to be a tempting solution, but this is well known to be problematic.
In \citet{Schreier20t} (Fig.\ 5) we have shown that Earth-like temperature profiles can be represented as a linear combination of some base vectors resulting from a singular value decomposition (SVD) of a large matrix comprising ``representative'' temperatures (comprising the 42 Garand atmospheres and the 11 M-Earth atmospheres introduced in subsection \ref{ssec:data}).
Hence we will use here a linear least squares fit to determine the expansion coefficients for the ``Chahine initial guess'' profile and then ``extrapolate'' this profile to the entire altitude range.
In addition this is used for interpolation to a dense altitude grid in the lower and mid atmosphere appropriate for radiative transfer modeling.
(For brevity this will be called ``extrapolation'' henceforth.)

Before starting the iterations according to \eqref{chahineRelax} one more issue has to be discussed: how to stop the process.
For nonlinear least squares solvers such as MINPACK \citep{More78} or NL2SOL \citep{Dennis81,Dennis81a} two convergence criteria are usually employed:
firstly the change of the estimated state vector (here temperature) is small
and the change of the residual norm (the norm of the model minus observed signal vector, here the radiance spectrum) is small (where ``small'' should be related to $S/N$).
Inspection of the radiance residual norm is clearly a natural choice for least squares.
Exploiting the deviation of the fitted temperature to the true temperature or the deviation of the model to the ``true'' spectrum is clearly impossible for analysis of true observations and is hence not used as a convergence criterions.

\begin{figure}
 \centering\includegraphics[width=\linewidth]{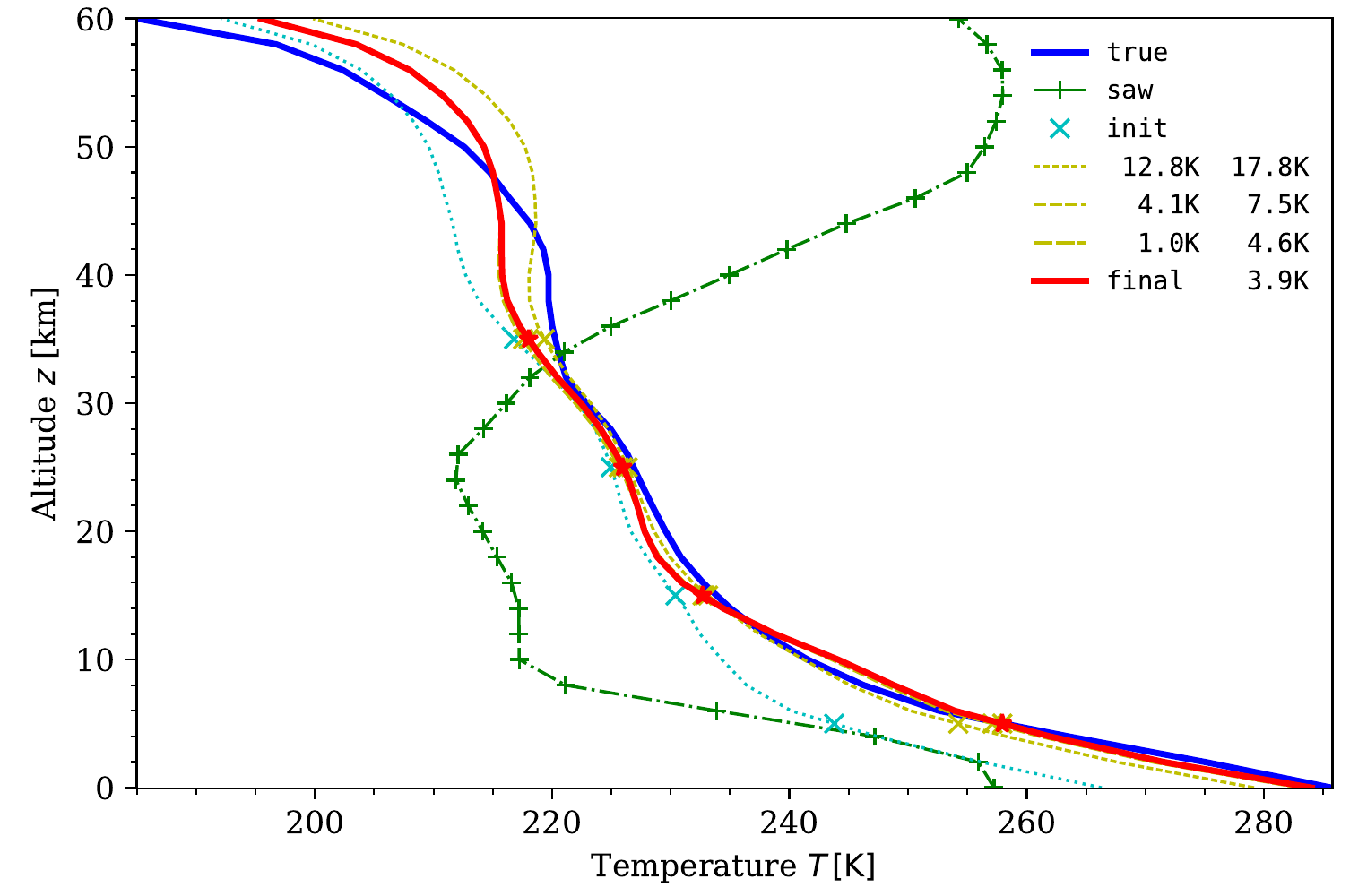}
 \caption{Chahine relaxation for AD\:Leo starting with SAW pressure, temperature (green), and concentrations ($S/N=100$, TIR-LW).
 The cyan crosses show the initial guess \eqref{equTempGuess} (similar to \qufig{secondGuess_adLeo}) and the cyan dotted line its extrapolation.
 Intermediate temperature profiles \eqref{chahineRelax} are shown in yellow, the numbers in the legend indicate the maximum temperature change ($T_{i+1}-T_i$) and maximum EBT deviation (observed - model).
 The red curve shows the final temperature profile with the maximum EBT deviation in the legend.}
 \label{chahineRelaxDemo}
\end{figure}

\begin{figure}
 \centering\includegraphics[width=\linewidth]{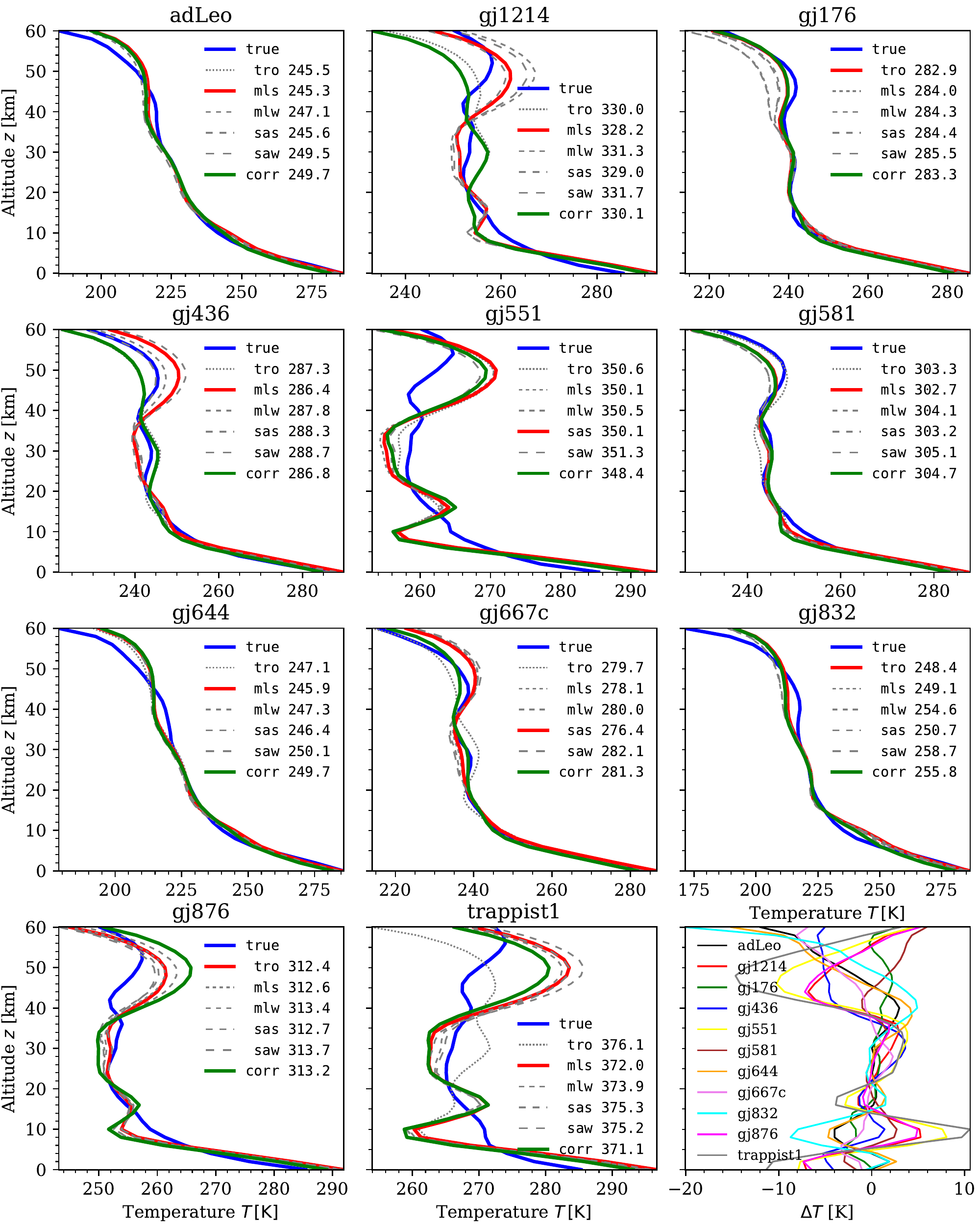}
 \caption{Comparison of atmospheric temperatures using Chahine relaxation for all Earth data (specified in the legend).
          The red solid line indicates the best fit according to the norm $\| \Delta I \|$ of the residual radiance spectrum (in units $\rm erg/s/(cm^2 \, sr \, cm^{-1})$, see legend).
          The green temperature shows the fit using the correct atmospheric densities.
          The last subplot shows the temperature differences (true - fit) for the best Earth model.
          (All $\nu \leftrightarrow z$ pairs with 5\,km tolerance, extrapolation with 4 base vectors, TIR-LW spectra as above.)
          }
 \label{chahineAllPairsAllMaps}
\end{figure}

For the analysis of the TIR-LW spectra using iterative Chahine relaxation \eqref{chahineRelax} we start with, e.g., Earth's SAW atmospheric data;
the SAW temperature is needed to compute the atmospheric transmission $\T$ and surface emission $I_\text{surf}$ in \EqRef{equTempGuess}.
The following stopping criteria have been used: a maximum change of the updated temperature less than 5\,K or a maximum change of the EBT of less than 5\,K.
For all M-Earths the relaxation stops after two or three iterations (see \qufig{chahineRelaxDemo} for an illustrative example).
The results (\qufig{chahineAllPairsAllMaps}) are consistent with those of \qufig{secondGuessAll_TIR2}:
for most planets the temperature is retrieved quite well for altitudes below about $40\rm\,km$, but GJ\:551 and Trappist-1 (and to a lesser extent GJ\:1214) appear to be problematic.
Nevertheless, for all planets the cold trap temperatures closely resemble the true one, and even for Trappist-1 the absolute temperature differences are small.

Further runs with other Earth model data (MLS, MLW, SAS, and tropical) essentially confirm these findings, see \qufig{chahineAllPairsAllMaps}.
For planets with very low minimum temperatures (AD\:Leo, GJ\:644, or GJ\:832) all models lead to almost identical temperatures with small differences only in the upper atmosphere.
Differences are clearly visible for GJ\:1214, GJ\:551, GJ\:876, and Trappist-1, i.e.\ planets with relatively high minimum temperatures.

Although the most favorable model is unknown for real observations we can nevertheless use the minimum radiance residuum norm $\| \Delta I \| = \| I_\text{obs}-I_\text{mod} \|$ as a hint for selection of the ``best'' fit.
In all but two cases this selected solution corresponds to the optimum solution according to the minimum norm of the EBT difference spectrum $ \Delta T_\text{B}$:
for GJ\:551 and GJ\:876 MLS give the smallest EBT difference, whereas SAS and TRO yield the smallest radiance difference (however, both norms are identical within four digits).
Using the mean EBT difference as criterion gives also different solutions for GJ\:581.

\begin{table}
\caption{Comparison of the radiance residuum norms (in radiance units $\rm erg/s/(cm^2\,sr\,cm^{-1})$) for iterative Chahine relaxation with the correct atmospheric pressure and concentrations
with fits using one of Earth's model atmospheres. The ``min'' and ``max'' columns give the range of residuum norms for
these fits. The second column lists the range of atmospheric temperatures, i.e.\ $\max(T)-\min(T)$ in Kelvin.}
\label{tab:normRad}
\begin{tabular}{lrrrrrrr}
\hline
            &           & \multicolumn{3}{c}{$S/N=10$}  &  \multicolumn{3}{c}{$S/N=100$}         \\
            & $\Delta T $ & corr. &   min   &   max   &  corr.   &   min  &   max  \\
\hline
AD\:Leo     &    100.6 & 249.7 & 245.3 & 249.5 & 42.43 & 41.2 & 49.2 \\
GJ1214      &     35.1 & 330.1 & 328.2 & 331.7 & 58.94 & 60.4 & 71.5 \\
GJ176       &     62.0 & 283.3 & 282.9 & 285.5 & 40.73 & 43.8 & 51.6 \\
GJ436       &     56.7 & 286.8 & 286.4 & 288.7 & 32.14 & 36.6 & 47.4 \\
GJ551       &     27.3 & 348.4 & 350.1 & 351.3 & 71.32 & 72.8 & 83.1 \\
GJ581       &     51.5 & 304.7 & 302.7 & 305.1 & 34.03 & 39.0 & 48.7 \\
GJ644       &    106.9 & 249.7 & 245.9 & 250.1 & 47.77 & 45.0 & 54.7 \\
GJ667c      &     69.8 & 281.3 & 276.4 & 282.1 & 36.33 & 34.8 & 48.5 \\
GJ832       &    114.2 & 255.8 & 248.4 & 258.7 & 60.31 & 47.8 & 67.6 \\
GJ876       &     35.2 & 313.2 & 312.4 & 313.7 & 55.08 & 57.8 & 67.7 \\
Trappist1   &     20.2 & 371.1 & 372.0 & 376.1 & 90.76 & 92.8 & 101.8 \\
\hline
\end{tabular}
\end{table}

In addition to the retrievals using the five Earth models \qufig{chahineAllPairsAllMaps} also shows the temperature fitted with the correct exoplanet pressure and composition
(clearly unknown in case of real observations); the initial guess temperature required to compute the weighting functions and total atmospheric transmission $\T(\nu,0)$ is set to the mean EBT.
Interestingly the norm of the radiance difference with the correct exoplanet composition is smaller than the norm with the best Earth composition only for some of the planets.
However, Table \ref{tab:normRad} indicates that the variability of these residual norms is always small, the difference of the largest and smallest norm is less than a few percent,
and the residual of all fits is very large because of the significant noise of the synthetic intensities (i.e.\ the residual is dominated by noise).

For synthetic measurements with less noise the residual norm of fits with different atmospheric models shows larger variations up to 41\% for $S/N=100$
(because of the smaller noise the stopping criterium has been tightened to 3\,K temperature change).
Note that the variation of the norms is especially large for planets with a large range of atmospheric temperatures.
Again the correct atmosphere does not always deliver the best fit, but the best Earth model and the correct densities always give similar temperature profiles.
Table \ref{tab:normRad} shows that the correct densities do not yield a model spectrum closer to the observed spectrum for the four M-Earths with the largest temperature gradients $\Delta T$ (AD\:Leo, GJ\:644, GJ\:667c, and GJ\:832).

At first glance this ``failure'' appears to be quite disturbing and we interpret this as follows:
The residual norm is clearly the essential number characterising least squares, i.e.\ the quantity to be minimised.
However, it is not the decisive quantity for Chahine relaxation; obviously the ratio of observed to model spectra in \eqref{chahineRelax} should finally be close to one, but there is no uniquely defined number for the progress of the relaxation and quality of the solution.

These results may also be considered as a hint that Chahine relaxation relies on a strong simplification, i.e.\ the intensity \eqref{schwarzschild} at a particular wavenumber can be approximated by the Planck emission at a particular altitude according to \eqref{schwarzschildApprox}.
This assumption clearly ignores the shape of the weighting functions (compare \qufig{wgtFct_tir2} bottom-left) and is apparently problematic when the temperature differences are large.
Furthermore, large temperature differences make the extrapolation more difficult.
Nevertheless, the last subplot of \qufig{chahineAllPairsAllMaps} demonstrates that in the mid atmosphere (about 15 to 35\,km) the temperature can be estimated within $\pm 5\rm\,K$.

\begin{figure}
 \centering\includegraphics[width=\linewidth]{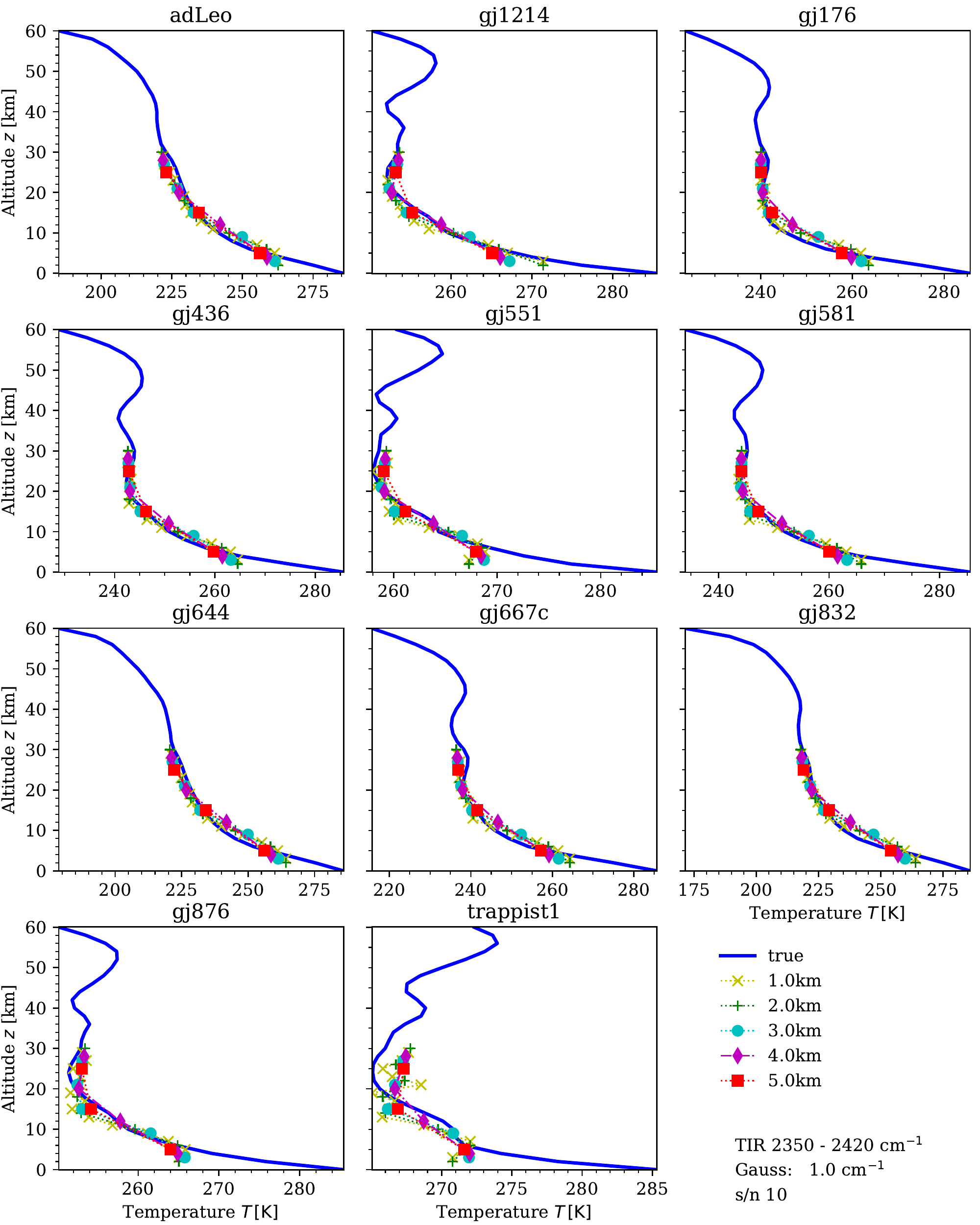}
 \caption{Estimate of atmospheric temperatures using all $\nu \leftrightarrow z$ pairs for MLS and TIR-SW (2350 -- 2420\cm, $S/N=10$, Gaussian with $\Gamma=1.0\rm\,cm^{-1}$.)}
 \label{secondGuessAll_TIR3}
\end{figure}

\subsection{TIR-SW} 
\label{ssec:tir3}
The shortwave TIR appears to be less favourable for temperature retrieval for several reasons:
The radiance values in the TIR-LW are higher compared to the TIR-SW (Fig.\ 11 in \citet[][]{Schreier20t}),
the TIR-LW is more favourable because of the higher star-planet contrast,
and the TIR-LW is also less affected by scattering.
Moreover, the TIR-SW weighting function peak heights do not cover altitudes above 30\,km (\qufig{wgtFct_maxima}),
and for shorter wavelengths reflection of thermal radiation at the surface is becoming increasingly important.
On the other hand, the TIR-SW weighting functions indicate some more sensitivity to the lowest atmosphere, and in fact IR instruments of meteorological satellites used for sounding of Earth's temperature \citep{Menzel18}, e.g.\ AIRS \citep{Chahine06} and IASI \citep{Hilton12}, exploit both regions.

\begin{figure}
 \centering\includegraphics[width=\linewidth]{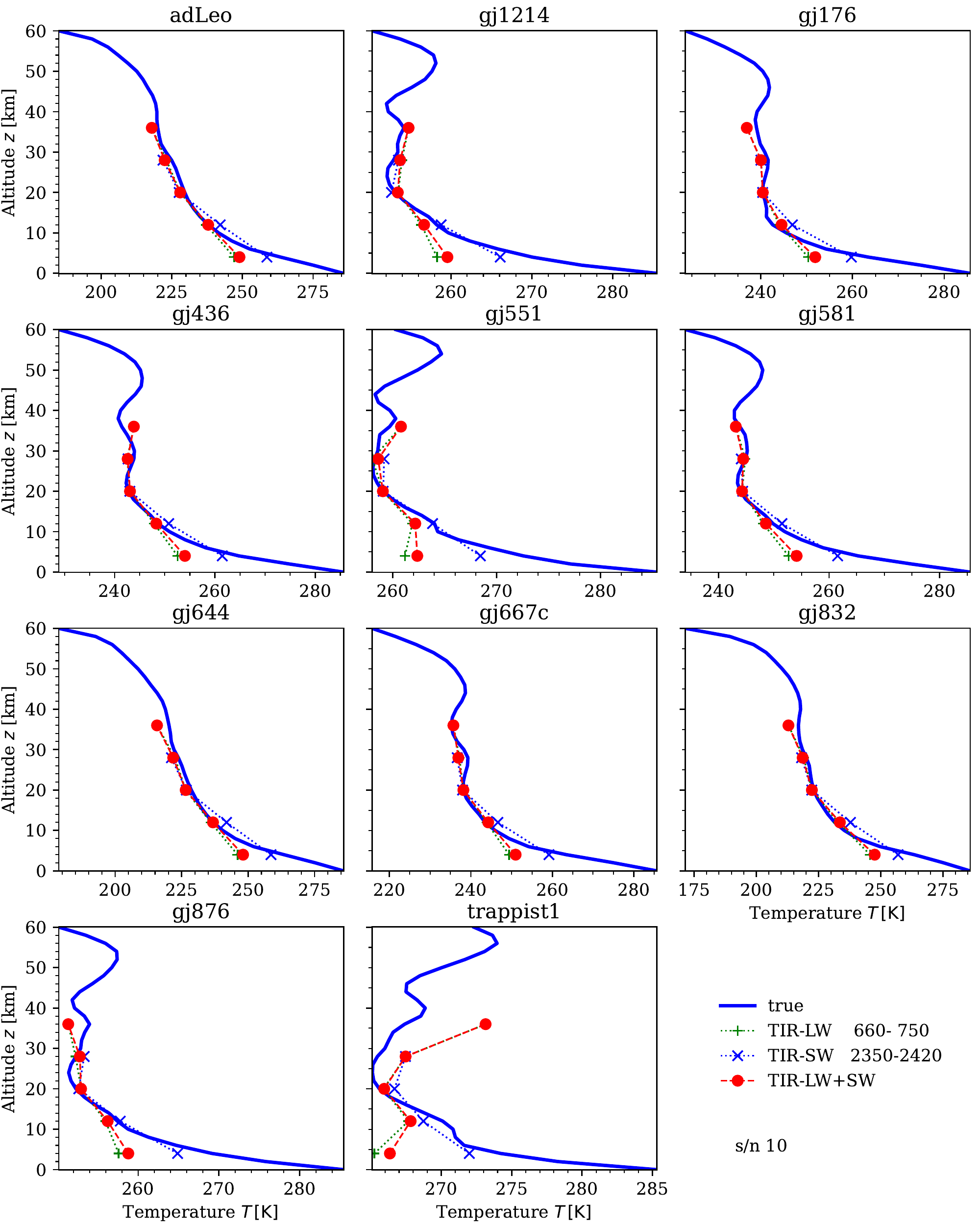}
 \caption{Comparison of atmospheric temperatures estimated from the TIR-SW, the TIR-LW, and the concatenated spectrum.
          (All $\nu \leftrightarrow z$ pairs with a 4\,km tolerance for MLS, $S/N=10$)}
 \label{secondGuessAll_TIR23}
\end{figure}

Figure \ref{secondGuessAll_TIR3} depicts the temperature inferred from the equivalent brightness temperatures \eqref{equTempGuess} in the TIR-SW using all $\nu \leftrightarrow z$ pairs for MLS.
Considering altitudes within one or two kilometers around a weighting function peak height gives zigzag temperature profiles as in the longwave analysis, cf.\ \qufig{secondGuessAll_TIR2}.
For the $5\rm\,km$ tolerance the profiles are smoother, however only four temperature values for the upper troposphere and lower stratosphere are estimated.
Compared to the TIR-LW estimates, \qufig{secondGuessAll_TIR2}, the profiles appear to be somewhat smoother and closer to the true temperature.
Trappist-1, however, is reasonable only in the $10\text{\,--\,}20\rm\,km$ range.
For other Earth mappings (MLW, SAS, SAW, tropical) temperatures are almost identical (not shown).

In \qufig{secondGuessAll_TIR23} results from temperature estimates using the SW and LW interval individually are compared with those from the combined spectrum (with 1722 data points, cf.\ \qufig{observedBT}).
The ''data fusion'' product clearly benefits from the sensitivity of the longwave spectrum beyond $30\rm\,km$;
however, it also inherits the underestimated temperature in the lowest atmospheric levels, especially for GJ\:1214, GJ\:551, GJ\:876, and Trappist-1.
A more sophisticated data fusion approach might possibly be able to avoid the shortcomings of the TIR-SW in the upper atmosphere and TIR-LW in the lower atmosphere.

\subsection{Trappist-1e} 
\label{ssec:Trappist1}

The Trappist-1 planetary system \citep{Gillon17t} orbiting a nearby M-dwarf has attracted considerable attention because several terrestrial-type exoplanets lie in the circumstellar habitable zone \citep[e.g.][]{Barstow16t,Grimm18,KrissansenTotton18,LustigYaeger19,Fauchez20t,KrissansenTotton22}.
Planets e and f have been the object of numerous studies \citep[e.g.][]{Barstow16t,Morley17jw,MikalEvans21}.
Recently, \citet{Wunderlich20} have used the newly developed radiation--convection--photochemistry model 1D-TERRA to study the feasibility of atmospheric characterisation,
in particular the possibility of finding any evidence for an ocean or biosphere.

Here we use these dry\&dead, wet\&dead, and wet\&live scenarios with varying \element{CO_2} levels (see subsection \ref{ssec:data}) for further tests of the Chahine methodology.
We generate synthetic observations in the same way as for the M-Earths (\qufig{observedBT}), i.e.\ monochromatic intensity spectra according to \eqref{schwarzschild} are convolved with a Gaussian and contaminated with noise.

\begin{figure}
 \centering\includegraphics[width=\linewidth]                                 {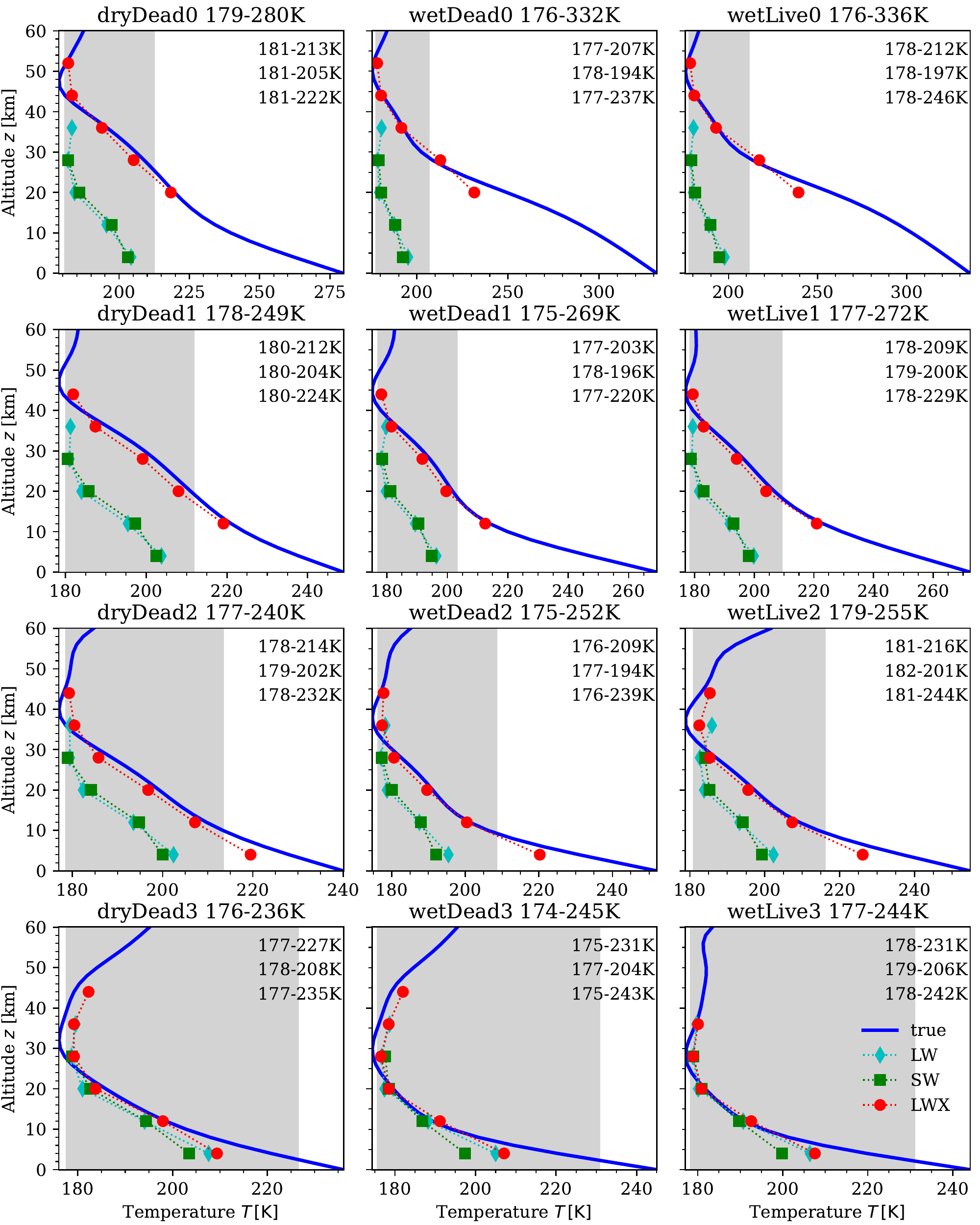}
 \caption{Trappist-1e temperatures from TIR-LW (660 -- 750\cm\ with $\Gamma=0.25\rm\,cm^{-1}$, cyan diamonds) and TIR-SW (2350 -- 2420\cm\ with $\Gamma=1.0\rm\,cm^{-1}$, green squares) with entire $\nu \leftrightarrow z$ MLS map.
          Third estimate with an extended TIR-LW (LWX, 660 -- 850\cm, red circles) and the correct map (i.e.\ correct \element{CO_2}). (All spectra with $S/N=10$).
          The number $n$ immediately following the planetary scenario description in each subplot title indicates the \element{CO_2} surface partial pressure, i.e.\ $p_\text{CO2} \approx 10^{-n}\rm\,bar$.
          (Regarding notation see subsection \ref{ssec:data}.)
          Minimum and maximum atmospheric temperatures are listed in the title, the range of the EBTs of the noise-free LW, SW, and LWX spectra are given inside the plot.
          The gray shaded area shows the EBT range for the SW and LW combination.}
 \label{TIR23_t1e_fullMapMLS}
\end{figure}

Figure \ref{TIR23_t1e_fullMapMLS} is similar to \qufig{secondGuessAll_TIR23} and shows atmospheric temperatures retrieved from the TIR-LW and TIR-SW equivalent brightness temperatures \eqref{equTempGuess} using the MLS weighting function peak altitudes  (cf.\ \qufig{wgtFct_maxima}) with a 4\,km tolerance (without iteration).
The LW and SW estimates are quite similar and relatively smooth.
For the \element{CO_2} $\text{VMR}=10^{-3}$ atmospheres (bottom row) the temperature near BoA is somewhat underestimated, but in the middle atmosphere up to about 40\,km the estimates are close to the truth.
However, for the atmospheres with larger \element{CO_2} concentrations the temperatures are clearly too cool.

\begin{figure}
 \centering\includegraphics[width=\linewidth]{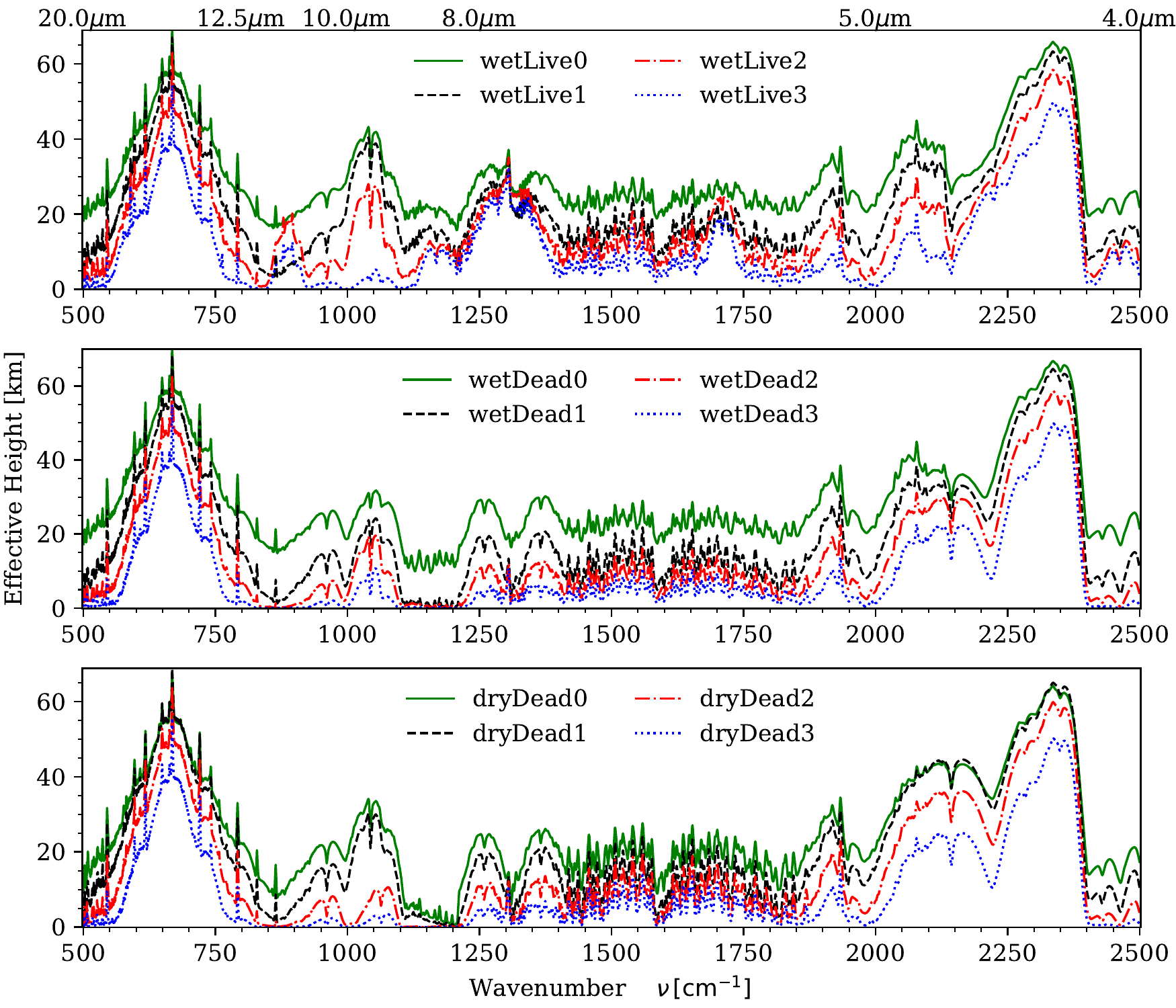}
 \caption{Trappist-1e effective heights for a Gaussian response function with $R=1000$. (Computed with GARLIC.)}
 \label{t1e_effHgt}
\end{figure}

Due to the large mixing ratios the atmospheres are optically thick in the spectral windows considered here, and radiation (photons) from the lower warm altitudes cannot propagate upwards to ToA (and the observer).
This interpretation is confirmed by the range of equivalent brightness temperatures given in the plot, i.e.\ the warm low atmosphere does not show up in the EBT spectra.
Analysis of effective height spectra (\qufig{t1e_effHgt}) that might be available from primary transit spectroscopy also indicates that the TIR-LW interval (660 -- 750\cm) is insensitive to the lower atmosphere.

\begin{figure}
 \centering\includegraphics[width=\linewidth]{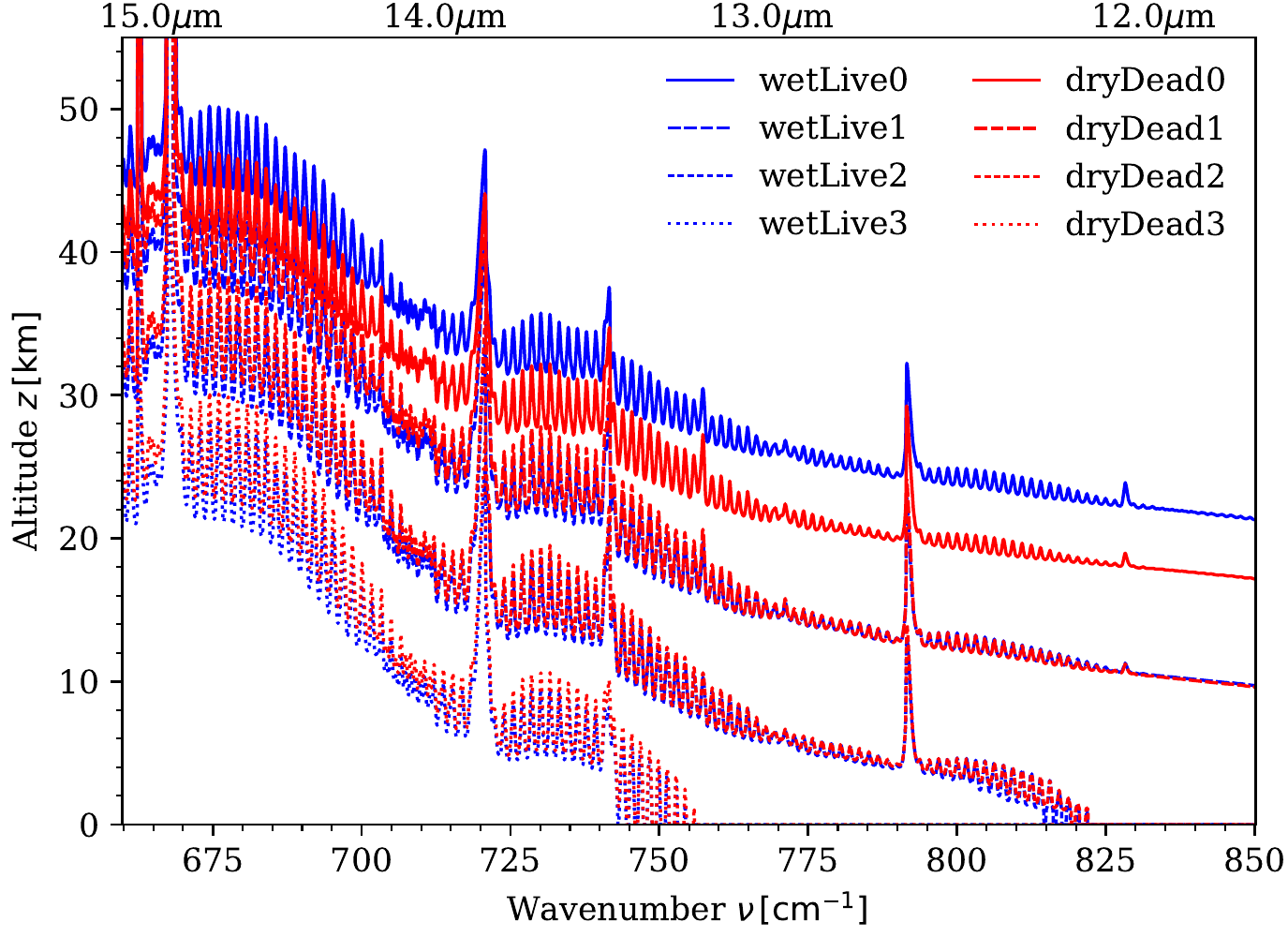}
 \caption{Trappist-1e altitudes of weighting function maxima in the extended LW-TIR (Gauss $\Gamma=0.25\rm\,cm^{-1}$). The wet\&dead weighting function maxima are largely similar to the wet\&live maxima.}
 \label{t1e_wgtFctMax}
\end{figure}

Furthermore, the success of the temperature estimates for the $\text{VMR}=10^{-3}$ cases suggests that the $\nu \leftrightarrow z$ mappings of Earth's atmospheres are not adequate for Trappist-1e atmospheres with more carbon dioxide.
\qufig{t1e_wgtFctMax} shows that with higher \element{CO_2} concentrations the location of the weighting function maxima move upwards by 5 or even $10\rm\,km$, whereas the ``nature'' of the planet (wet vs.\ dry, dead vs.\ alive) does not have a big impact.

In summary, Figures \ref{t1e_effHgt} and \ref{t1e_wgtFctMax} suggest that atmospheres with abundant \element{CO_2} are not well reproduced from TIR-LW and TIR-SW spectra especially in the lower regions (\qufig{TIR23_t1e_fullMapMLS}).
Therefore we also considered an extended spectral range along with the correct weighting function peak heights (i.e.\ weighting functions in the extended TIR-LW computed with the true concentrations) which considerably improved the estimate (esp.\ in the mid atmosphere, \qufig{TIR23_t1e_fullMapMLS} red dots; however, the sensitivity to the lower atmosphere is lost, i.e.\ no mappings with altitudes below 10 or $20\rm\,km$ for the 10\% and 50\% \element{CO_2} planets).

Having demonstrated that atmospheric temperatures can be estimated from equivalent brightness temperatures \eqref{equTempGuess} using an extended TIR-LW interval if the composition is known, we now examine iterative Chahine relaxation \eqref{chahineRelax}.
The assumption of known abundances appears reasonable since some objects will have information from transmission spectra.
In particular quantifying \element{CO_2} might be possible due to e.g.\ its dominant absorption bands
(clearly visible around 660\cm\ and 2400\cm\ in \qufig{t1e_effHgt}); concentration estimates for other gases however might be more difficult.

\begin{figure}
 \centering\includegraphics[width=\linewidth]                                 {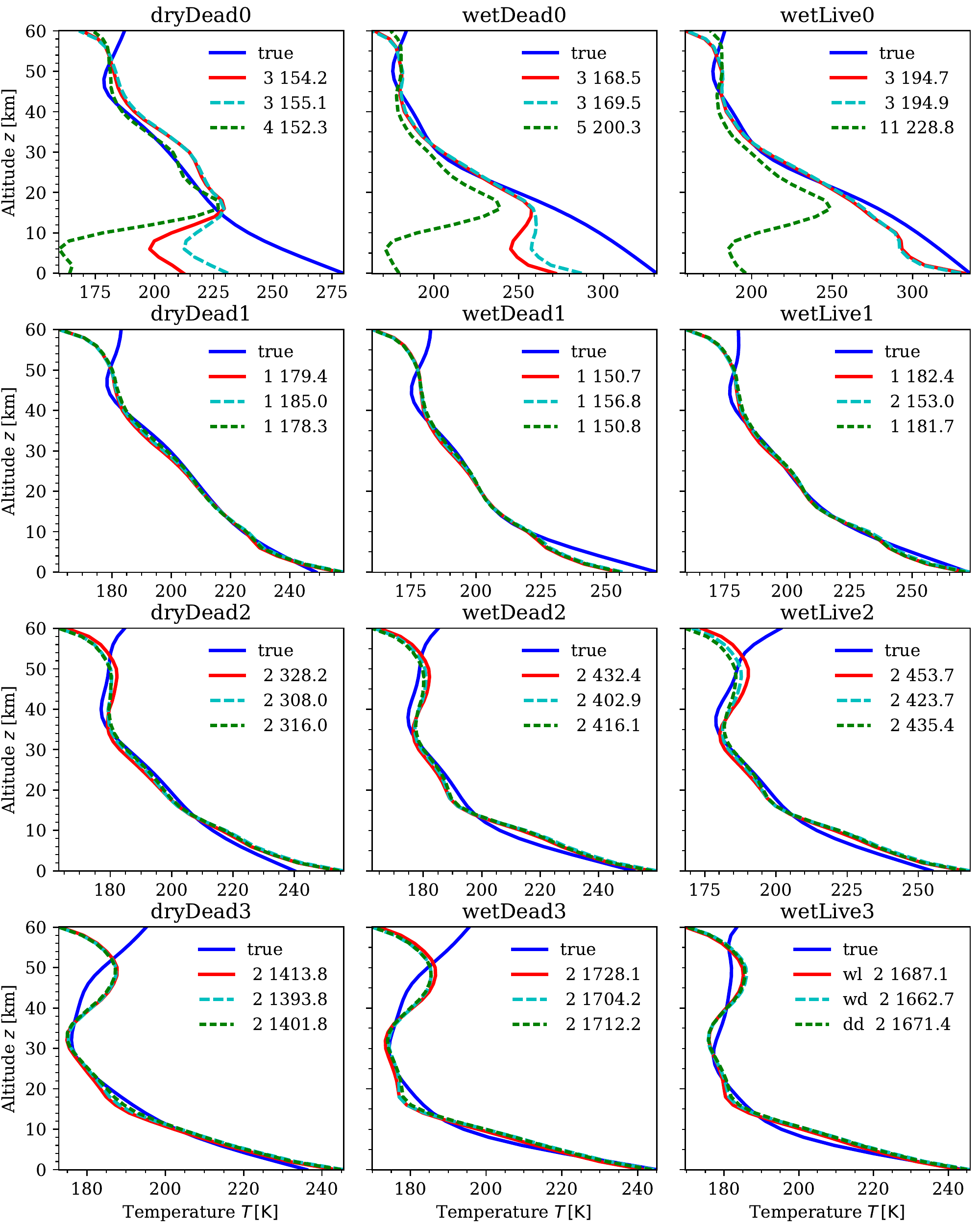}
 \caption{Chahine iterative estimate of Trappist-1e temperatures with correct \element{CO_2} abundance and varying scenarios:
          wet\,\&\,live (red), wet\,\&\,dead (cyan long dashed), and dry\,\&\,dead (green dashed).
          SVD ``extrapolation'' with four base vectors.
          Extended TIR-LW spectra as above. The legend lists the number of iterations and the radiance residual norm.}
 \label{t1e_TIR2x_iterative_mls}
\end{figure}

First we examine whether we can then distinguish the temperature profiles for the three scenarios (wet vs.\ dry, dead vs.\ live), i.e.\  we compare retrievals using different surface and lower atmosphere scenarios
(for example, for the wet\&live planet ``observation'' (right column in \qufig{t1e_TIR2x_iterative_mls}) we also investigate fits using boundary parameters of the dead scenarios with the correct \element{CO_2}).

Figure \ref{t1e_TIR2x_iterative_mls} (top) shows that the effect of changing the scenario is largest for the high \element{CO_2} abundance:
in the lower atmosphere temperature is significantly underestimated for the two dead scenarios
which is likely related to the fact that the effective heights (\qufig{t1e_effHgt}) and weighting functions (\qufig{t1e_wgtFctMax}) do not show any sensitivity to the lowest atmosphere (esp.\ for the wet cases).
For the observation of the wet\&live $\text{VMR}\!=\!0.5$ planet (top right) both wet models yield an only moderately underestimated temperature.
The difficulty of the high \element{CO_2} atmospheres is also demonstrated by the increased number of iterations required for convergence.
Fits with large residuum norm show the largest deviations between true and retrieved temperature in the mid atmosphere.
For moderate and low \element{CO_2} (other rows in \qufig{t1e_TIR2x_iterative_mls}) the impact of the scenarios is only weak, and only one or two iterations are necessary.
Results here are close to the true profile (blue line) in the low and mid atmosphere, although deviations are clearly evident in the upper atmosphere where the weighting function exhibits little sensitivity.

\begin{figure}
 \centering\includegraphics[width=\linewidth]                                 {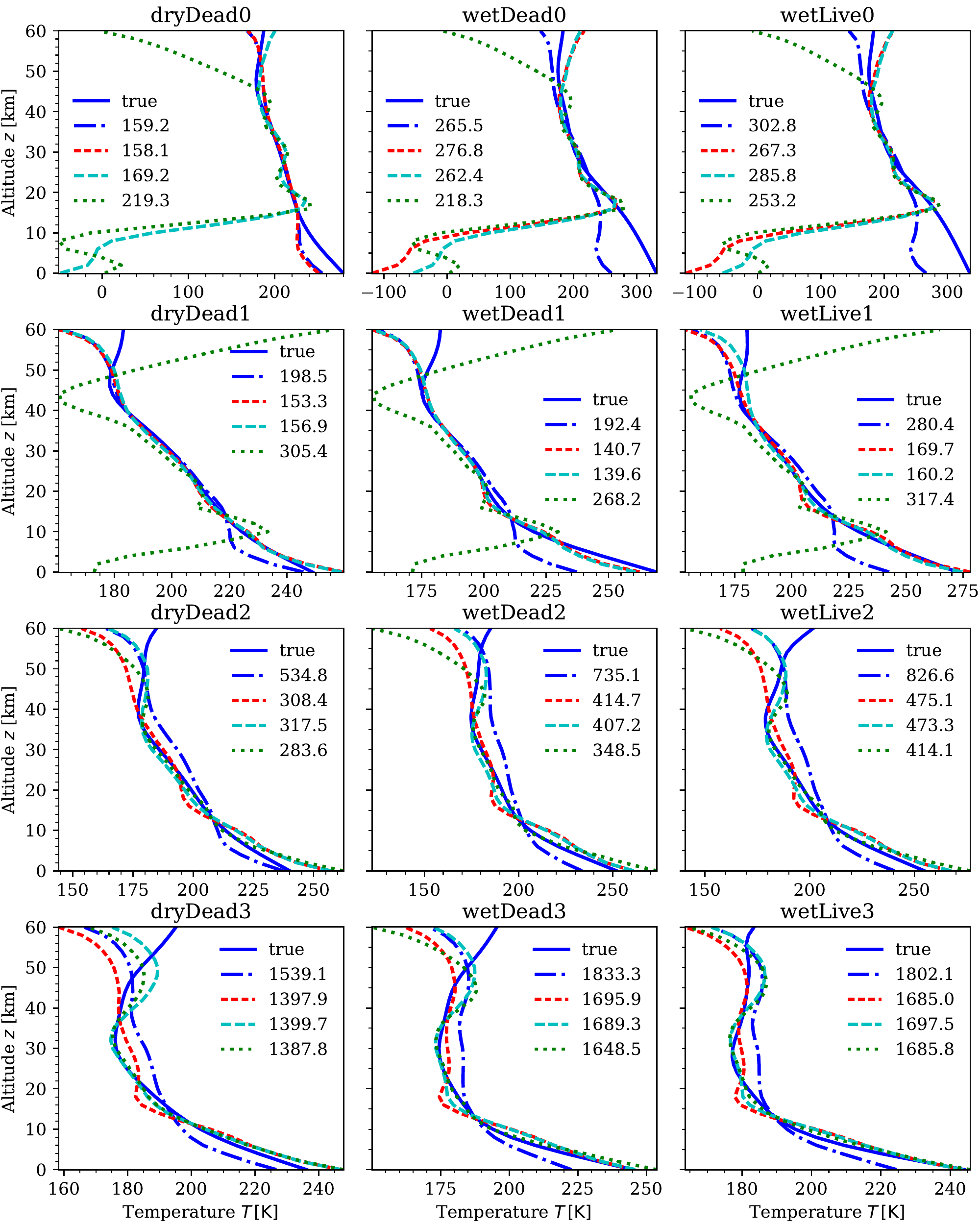}
 \caption{Chahine iterative estimate of Trappist-1e temperatures starting with MLS $\nu \leftrightarrow z$ mapping and molecular profiles with correct column density.
          The retrieved temperatures correspond to SVD ``extrapolation'' with two (blue dash-dotted), three (red dashed), four (cyan long dashed), and five (green dotted) base vectors. (Numbers in legend list the radiance residual norm, spectra as above.)}
 \label{t1e_TIR2x_iterative_iso}
\end{figure}

A second set of runs has been conducted assuming MLS molecular profiles scaled to the correct Trappist-1e column density, see \qufig{t1e_TIR2x_iterative_iso}.
Similar to Earth the Trappist-1e atmospheres have \element{CO_2} VMR profiles almost constant in altitude, and the \element{H_2O} VMRs are strongly decreasing for pressures below $100\rm\,mb$ \citep[see Fig.\ 7 in][]{Wunderlich20}. 
(In contrast the \element{CO_2} VMR of all M-Earths increases by 10 -- 20\% for $p<100\rm\,mb$ \citep[see Fig.\ 4 in][]{Wunderlich19}.)
However, at least for Earth, the total optical depth is essentially dominated by the \element{CO_2} contribution \citep[see Fig.\ A3 in][]{Schreier20t},
which suggests that for these Trappist-1e atmospheres with even higher \element{CO_2} concentrations this dominance will be even stronger.
Hence, the assumption of isoprofiles does not have a strong impact on the quality of the retrievals, i.e.\ temperature in the mid atmosphere can be estimated with little deviation from the truth.

Figure \ref{t1e_TIR2x_iterative_iso} also shows the importance of the ``extrapolation'' scheme based on the expansion using singular vectors.
Although a temperature profile representation using only two base vectors delivers temperatures close to the true even for the lowest atmosphere (including the $\text{VMR}=0.5 \cdot 10^0$ cases, top row),
a profile expansion with three or four base vectors appears to be more reasonable.
Using five base vectors works well for $\text{VMR} \le 10^{-2}$, but clearly fails for very high \element{CO_2} concentrations.
The radiance residual norms shown in the plot also indicate that three or four base vectors can be used reliably except for the high \element{CO_2} concentrations.

Further tests have also been conducted with reduced or increased \element{CO_2} concentrations.
For moderate changes with quarter, half, double and quadruple isoprofiles the retrieved temperature profiles are still close to the true temperature.
(See the further discussion in subsection \ref{ssec:discAtm}.)


\section{Discussion}
\label{sec:disc}

\subsection{Spectral resolution}
\label{ssec:resolution}

The retrievals reported so far have been conducted assuming moderate resolution TIR-LW and TIR-SW spectra.
The motivation for $R>2500$ has been discussed in \citet[][Figure 1]{Schreier20t} where we showed that for decreasing resolution the sensitivity to upper atmospheric layers decreases:
for a Gaussian response function with HWHM $\Gamma=0.25\cm$ (corresponding to a resolution $R=2800$ at $700\cm$ (wavelength $14.3\mue$)) the maximum of the weighting function peak height spectrum (cf.\ \qufig{wgtFct_maxima}) reaches altitudes above $40\rm\,km$;
a coarser resolution reduces the peak height (for $\Gamma=1.0\cm$ ($R=700$), $\Gamma=2.0\cm$ ($R=350$), and $\Gamma=7.0\cm$ ($R=100$) the maxima lie at 40\,km, 24\,km, and 18\,km, respectively).

The TIR-LW and TIR-SW \element{CO_2} bands can be observed by the Medium Resolution Spectrometer (MRS) of the JWST Mid Infrared Instrument (MIRI) with a resolving power $R$ of about 2500 \citep{Rieke15}
(the Low Resolution Spectrometer (LRS) only sees the TIR-SW band with a resolution $R=100$).
However, \citet{Morley17jw} caution that ``for temperate planets spectroscopy with JWST-MIRI will likely be unrealistically expensive''.
In the JWST Guaranteed Time Observations (GTO) program several exoplanets have already been observed with the Near Infrared Imager and Slitless Spectrograph (NIRISS), NIR Spectrograph (NIRSPEC), and NIR Camera (NIRCam) instruments, that can deliver spectra with wavelengths up to $5\rm\,\mu m$ at low and medium resolution.

\begin{figure}
 \centering\includegraphics[width=\linewidth]{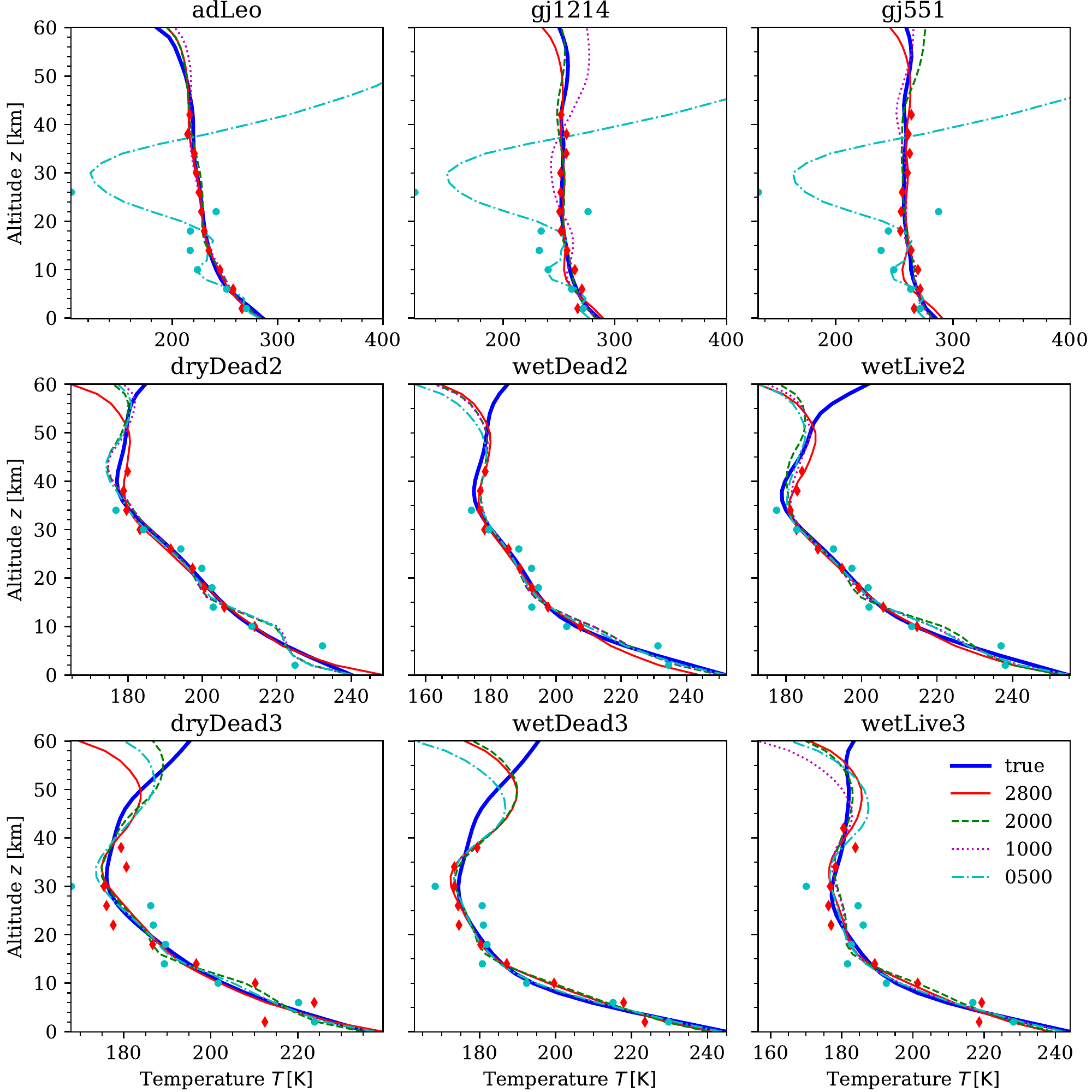}
 \caption{Impact of resolution on Chahine iterative temperatures estimates. MLS initial guess atmosphere, SVD ``extrapolation'' with 4 base vectors.
 The red diamonds and cyan circles show the final update according to the relaxation equation \eqref{chahineRelax} for $R=2800$ and $R=500$, respectively. TIR-LW (M-Earths) or TIR-LWX (Trappist-1e).
 The solid, dashed etc.\ lines show the corresponding extrapolated temperature profiles.}
 \label{TIR2-resolution}
\end{figure}

For an assessment of the impact of resolution on the retrieval, synthetic observations have been generated with different resolutions for selected exoplanets.
Noise has been adjusted assuming a square root relationship between resolution and $S/N$.
The results depicted in \qufig{TIR2-resolution} confirm the expectations discussed above, i.e.\ with decreasing resolution information on the upper atmosphere is diminishing.
In particular, for $R=500$ Chahine estimates according to \eqref{chahineRelax} are only available for altitudes below $20\rm\,km$ (shown as cyan circles) and ``extrapolation'' is clearly problematic.

\subsection{Surface emission}
\label{ssec:surface}

Apart from the type of inversion (least squares vs.\ Chahine relaxation) the methodology used here is largely identical to that of our previous feasibility study.
However, there is one large difference worth discussing: in \citet{Schreier20t} we have ignored surface emission.
Obviously the first term in \eqref{schwarzschild} is mandatory for modeling IR spectra in atmospheric window regions where the atmosphere is relatively transparent such as the $ 800 \text{\,--\,} 1200 \rm\,cm^{-1}$ interval for Earth (except for \element{O_3} absorption around $9.6\rm\,\mu m$).
For the terrestrial exoplanets, transmission $\T \approx 0$ in the \element{CO_2} bands considered in \citet{Schreier20t} (LW: $660 \text{\,--\,} 720 \rm\,cm^{-1}$), hence the surface contribution has no impact on mid atmospheric temperature estimates.

However, for wavenumbers beyond $720 \rm\,cm^{-1}$ corrections cannot be neglected, so an estimate of the surface temperature is required.
If wavenumbers with transmission close to one are observed, the corresponding radiance can be used to estimate $T_\text{surf}$.
Here we approximate surface temperature by the largest equivalent brightness temperature observed, assuming that the maximum atmospheric temperature corresponds to the BoA temperature and is approximately equal to the surface temperature.
(Although warmer temperatures are possible at the stratopause, these do not contribute strongly to the spectrum as indicated by the weighting functions (cf.\ \qufig{wgtFct_tir2})).

Due however to noise, the largest EBT may not always be the best estimate of the surface temperature.
Alternatively the maximum of a smoothed EBT spectrum (e.g.\ by running averages) or the mean of the largest ten etc.\ values could be used.

\subsection{Molecular spectroscopy data}
\label{ssec:discHit}
Similar to \citet{Schreier20t} we have used the very first edition of the \textsc{Hitran} database \citep{Rothman87} rather than the latest version to speed up the computations.
This simplification appears to be justified for the feasibility study presented here, but is clearly inadequate for analysis of real observations.
In our previous study we discussed the then latest 2016 version \citep{Gordon17}, here we provide a brief update for the current \textsc{Hitran} 2020 \citep{Gordon22}.

For a realistic modeling of IR emission spectra in the $ 660 \text{\,--\,} 750 \rm\,cm^{-1}$ interval line data in an enlarged interval have to be considered to properly account for line wing contributions and convolution with the spectral response function.
Hence lines in the range $ 648.75 \text{\,--\,} 761.25 \rm\,cm^{-1}$ are read:
\textsc{Hitran} 2020 knows 449445 lines of 29 molecules, which reduces to 79394 lines of the five main IR absorbers (\element{CO_2}, \element{O_3}, \element{N_2O}, \element{CH_4}, \element{H_2O}).
In contrast, \textsc{Hitran} 86 returns 16003 lines of 4 molecules;  furthermore removing weak lines results in 4255 lines total including 2827 lines of carbon dioxide.
Despite this drastic reduction of active lines the final spectra do not change significantly:
For the MLS atmosphere the peak height of the weighting function is modified by a few hundred meters only (except for $\Delta z \approx 0.8\rm\,km$ at the high wavenumber end of the interval);
Likewise, the difference of the equivalent brightness temperature spectra is usually less than $1\rm\,K$ (with a maximum $|\Delta T_\text{B}| \approx 1.8\rm\,K$ at $718\rm\,cm^{-1}$).

Air-broadened half-widths are listed in the \textsc{Hitran} database since its beginnings half a century ago, whereas self-broadening was introduced later with the 1986 release \citep{Rothman87}.
Collision (pressure) broadening parameters for perturbers other than ``air'' (i.e.\ Earth's \element{N_2}, \element{O_2}) can be important for modeling spectra of other planets and were included only recently in \textsc{Hitran} \citep{Gordon17}.
For the terrestrial \element{N_2-O_2} dominated atmospheres considered here the \textsc{Hitran} 86 broadening parameters are clearly appropriate (with the possible exception of the 50\% \element{CO_2} Trappist-1e scenarios).
In this context it might also be worth noting that the line strengths in \textsc{Hitran} are also tuned to Earth, i.e.\ the strengths are scaled by the relative natural abundance of the isotopes in Earth's atmosphere.%
\footnote{\url{https://hitran.org/docs/iso-meta/}}

In addition to line transitions continua can be also important, especially water self and foreign continua \citep{Shine12} and collision induced absorption \citep[CIA, ][]{Richard12,Karman19}.
Comparison of EBT spectra (MLS atmosphere, $R=100$) computed with and without \element{H_2O}, \element{CO_2}, \element{O_2} and \element{N_2} corrections \citep{Clough89} indicates negligible differences in the TIR-LW regions considered here;
however, in the TIR-SW differences up to almost 8\,K show up around $2400 \rm\, cm^{-1}$ in the right wing (short wavelength) of the \element{CO_2} band.
(For the entire TIR ($500 \,\text{--}\, 2500 \rm\, cm^{-1}$) maximum differences up to $\approx 10\rm\,K$ at $1600 \rm\, cm^{-1}$ can be seen.)

\subsection{ToA 60 vs 120\,km}
In \citet{Schreier20t} all runs have been performed with a ToA at 60\,km, mainly because the \citet{Garand01} atmospheres are defined only for pressures down to $0.08\rm\,mb$ (about 60 to 65\,km), but also to speed up the computations.
Likewise, the M-Earth atmospheres of \citet{Wunderlich19} have ToA altitudes in the range 61 to 74\,km.
However, for IR radiative transfer modeling altitudes up to about 100\,km might be important (the AFGL data \citep{Anderson86} are given up to 120\,km).
For an assessment of the importance of the upper atmosphere we have compared radiances for 60 and 120\,km ToA and the MLS atmosphere.
In the TIR-LW region the EBT difference is usually less than 1\,K except for the strong radiance peak at 668\cm\ and in the right wings of the \element{CO_2} $\nu_2$ band.

\subsection{Geometry}
Observations of exoplanet thermal emission will deliver disk averaged spectra only, whereas we have used a single line-of-sight assuming a strict nadir view.
A common approximation for disk averaged spectra is to model a slant path with about $35^\circ$ from nadir.
The equivalent brightness spectra for the vertical and slant paths differ by less than 2\,K in the center of the TIR-LW band (sensitive to the mid atmosphere) with somewhat larger differences for $\nu > 700 \rm\,cm^{-1}$.
On the other hand, the peak altitudes of the weighting functions are shifted downwards by some 10\,km (see \qufig{wgtFct_tir2} lower left panel).
As a consequence, the magnitude of the retrieved temperature will not change significantly, but the associated altitudes could change by several kilometers.
It is also important to note that the preliminary constraints discussed in subsection \ref{ssec:prelim} are independent of the viewing angle.

\subsection{Auxiliary data, initial guess and a priori}
For the solution of ``real world'' inverse problems a large variety of auxiliary data are required, e.g.\ instrument parameters or observation geometry.
In the case of atmospheric IR spectroscopy these auxiliary data also comprise molecular optical properties (e.g.\ line lists or k-distributions);
for temperature sounding these data also include pressure and molecular concentrations.
These auxiliary data are often denoted ``a priori''.

However, for optimal estimation \citep{Rodgers76,Rodgers00} this solely refers to the knowledge of the unknown state vector (e.g.\ temperature) prior to any measurement
(all other auxiliary parameters are treated as ``model parameters'').
OE provides an estimate close to the a priori state vector where the relative weight of observation and a priori is determined by the respective measurement and a priori covariance matrices.
(Note that this weighting is independent of the weighting function $\upartial \T / \partial z$ defined in subsection \ref{ssec:irrt}.)
Obviously a priori knowledge of exoplanetary atmospheric properties is scarce (see e.g.\ \citealt{Shulyak19} for a thorough discussion).
\citet{Barstow20c} noted that ``the dependence of OE on an informative prior means that it is less appropriate for exoplanets'', which motivated the upgrade of the NEMESIS code \citep{Irwin08} with MultiNest \citep{Feroz09}.
Monte Carlo type retrievals are less sensitive to a priori, but for robust retrievals thousands to millions of time expensive forward model evaluations are required \citep{Fortney21}.

Some prior idea of the atmospheric state is required for Chahine-type inversions to compute the weighting functions and the total atmospheric transmission attenuating the surface emission (cf.\ Eqs.\ \eqref{schwarzschild} and \eqref{equTempGuess}, \eqref{chahineRelax}).
However, our simulations have demonstrated that the inferred M-Earth temperatures are largely independent of the assumed Earth atmospheric model and/or initial guess temperature profile.

Parameterisations as proposed by \citet{Madhusudhan09,Paris13r,Morley17gj} or \citet{Fossati20} are frequently used for temperature retrievals \citep[for a detailed discussion see Section 3 of][]{BarstowHeng20}.
Obviously, these parameterisations are motivated by physical insight and can also be considered as a kind of a priori knowledge.

For the iterative solution of nonlinear inverse problems by least squares \citep[e.g.][]{Schreier20t}, OE \citep[e.g.][]{Irwin08} (technically a constrained least squares) or Chahine relaxation, an initial guess is mandatory.
For the Earth-like exoplanets considered here a climatological Earth temperature profile or an isoprofile defined by the mean of the observed EBT can be used to start the iteration.
Alternatively, an atmospheric characterisation resulting from coupled photochemistry-climate codes such as 1D-TERRA \citep{Wunderlich20} can be used.

\begin{figure}
 \centering\includegraphics[width=\linewidth]{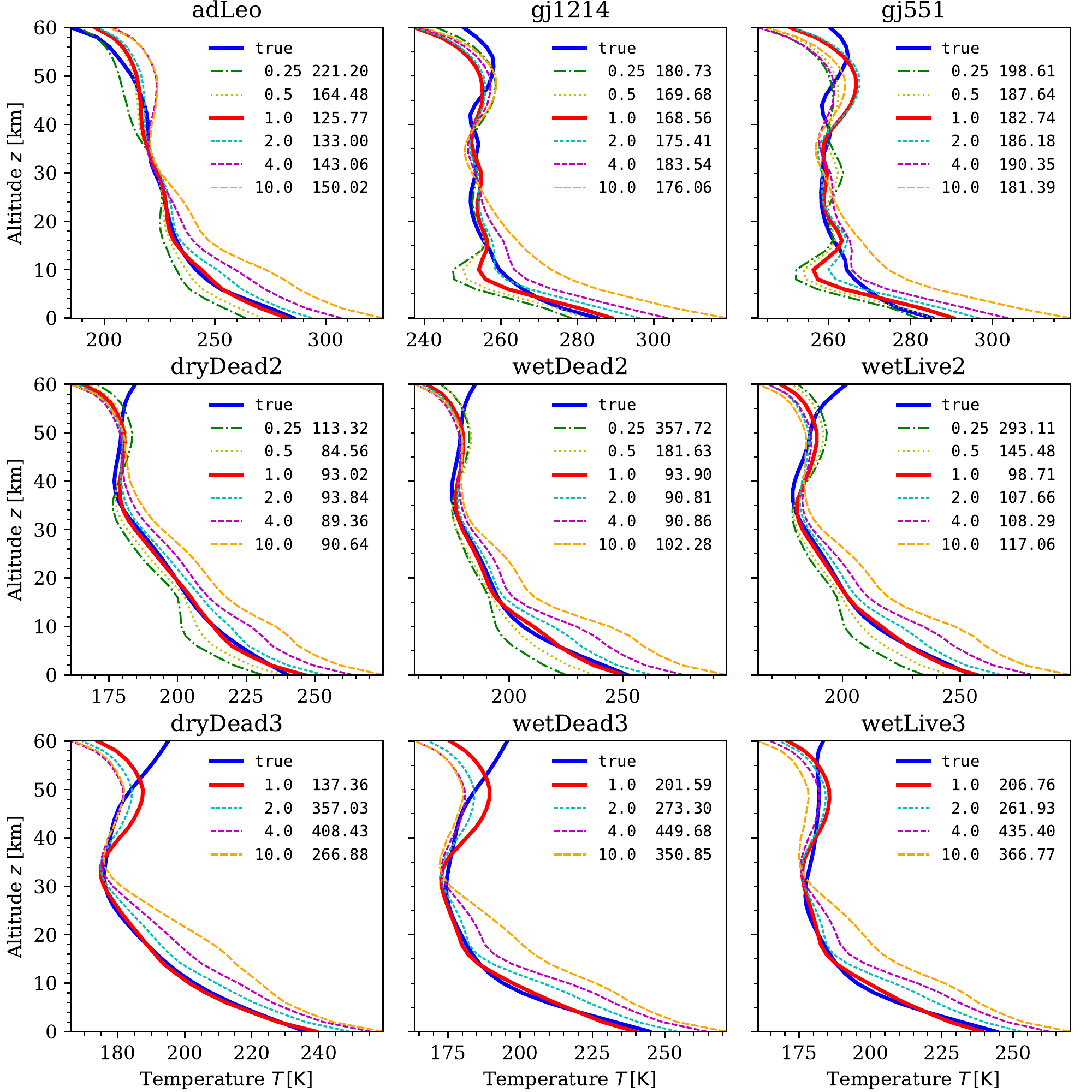}
 \caption{Impact of \element{CO_2} concentration. MLS initial temperature, SVD ``extrapolation'' with 4 base vectors, $S/N=20$, TIR-LW (M-Earths) or TIR-LWX (Trappist-1e) with Gaussian $0.25\rm\,cm^{-1}$.
 The legend indicates the factor used to scale the \element{CO_2} VMR isoprofile of the model atmosphere and the radiance residual norm.}
 \label{TIR2-CO2}
\end{figure}

\subsection{Atmospheric composition}
\label{ssec:discAtm}
Thermal emission IR spectra are sensitive to temperature but provide little information on atmospheric (molecular) composition.
However these data are mandatory for radiative transfer modeling.
In particular knowledge of the \element{CO_2} concentration is important for temperature sounding exploiting its strong TIR-LW and TIR-SW bands.
Actually the concentrations of all molecules absorbing in the spectral region to be analysed are required:
in the case of the two TIR bands, \element{H_2O}, ozone (\element{O_3}, longwave only) and methane (\element{CH_4}, shortwave) absorption are relevant.
Note that \element{H_2O} does not have a large impact on the TIR-LW weighting functions.

Assuming \element{CO_2} concentrations which are too low or too high likely has an impact on the quality of the retrievals.
Further test runs have been performed with a setup similar to \qufig{t1e_TIR2x_iterative_iso}, but with \element{CO_2} isoprofiles scaled by factors from 0.25 up to 10 in the model atmosphere.
Figure \ref{TIR2-CO2} demonstrates that in most cases the residual norm is larger for incorrect \element{CO_2} mixing ratios;
in some cases the iteration fails (e.g.\ for the 0.1\% \element{CO_2} Trappist-1e atmospheres, usually because of negative temperatures), the number of iterations becomes larger, or the temperature profile shows stronger oscillations
(zigzag profiles are unrealistic because the observed radiance is an integral \eqref{schwarzschild} that is insensitive to small-scale ``perturbations'' of temperature and/or molecular densities).

An independent estimate of concentrations is therefore desirable.
Carbon dioxide has several strong bands throughout the IR:
in addition to the TIR bands there are further strong rotation-vibration bands in the shortwave and near IR (SWIR, NIR) at $2.7\rm\,\mu m$, $2.0\rm\,\mu m$, and $1.6\rm\,\mu m$. 
In fact the NIR bands as well as the TIR-SW band enabled the identification of \element{CO_2} in the atmosphere of the Saturn-mass exoplanet WASP-39b with various JWST instruments \citep{Ahrer23,Alderson23,Rustamkulov23} and
the latter two SWIR bands are used operationally for monitoring of Earth's \element{CO_2} budget by several satellite missions such as OCO-2/3 or GOSAT \citep{Crisp04,Kuze09}.
Regarding exoplanets the analysis of effective height / transit depth spectra provided by primary transit observations is therefore valuable. 
Moreover, the ratio of observed signals in two appropriate filter bands is a sensitive indicator of  \element{CO_2} atmospheric concentration \citep{Rieke15}.
For the joint analysis of primary and secondary transits see also \citet{Griffith14}.


\section{Summary and Conclusions}
\label{sec:concl}

Temperature profiles of Earth-like exoplanets orbiting M-dwarfs have been retrieved from synthetic thermal IR emission spectra using Chahine-type relaxation methods along with a line-by-line radiative transfer code.
The essential assumption is that for a particular wavenumber the outgoing radiation arises from a corresponding, well-defined altitude:
In the band centre absorption is strong and only radiation from the upper atmosphere will be seen remotely;
in the band wings absorption is weak and even photons from the lower atmosphere can traverse the entire atmosphere to the ToA and beyond.
The feasibility of this method has been demonstrated using synthetic noise-contaminated observations of various Earth-like planets with \element{N_2}-\element{O_2} dominated atmospheres orbiting M-dwarfs and for Trappist-1e planets of different surface conditions (wet/dry and dead/alive) and different carbon dioxide concentrations up to about 50\%.
(Note that the assumption of \element{N_2}-\element{O_2} dominance is questionable for the high \element{CO_2} Trappist-1e scenarios.)

The equivalent brightness temperature (EBT) spectrum corresponding to the observed intensity can be used to deliver temperature estimates extremely quickly:
The minimum and maximum EBT values provide first constraints on the range of atmospheric temperatures independent of any a priori knowledge.
The EBT in the carbon dioxide absorption bands can be readily ``translated'' (within seconds) to mid atmospheric temperatures.
Using a handful of manually selected data points is problematic because of the noise, hence averaging of some neighboring spectral pixels is required (cf.\ \qufig{firstGuess8adLeo}).
However, exploiting the entire intensity spectrum and the corresponding wavenumber-altitude mapping, as defined by the weighting functions, is preferably and clearly advantageous (\qufig{secondGuess_adLeo}).
Furthermore, using both TIR-LW and TIR-SW data can be helpful to overcome the limited altitude sensitivity range of one region alone (\qufig{secondGuessAll_TIR23}).
In any case, the quality of this guess is however related to an appropriate knowledge of the \element{CO_2} concentrations, in particular in the case of the \element{CO_2}-rich Trappist-1e planets.

Iterative relaxation allows a refinement of the first guess, however, the success of the improvement relies on the inter/extrapo\-lation used to complete the limited set of $T$ data.
Note that an update of the temperature will change cross sections, transmission, and weighting functions and hence also the wavenumber-altitude mapping.

Compared to classical nonlinear least squares fitting a clear advantage of the Chahine relaxation (or the related Smith and Twomey schemes) is the fact that no Jacobian (derivatives of the intensity $I$ w.r.t.\ to the state vector elements) are required.
This is clearly beneficial when finite differences are used to approximate the Jacobian as for (nonlinear) optimal estimation or in the SVEEEETIES study \citep{Schreier20t} using Py4CAtS
(GARLIC exploits algorithmic differentiation, where the overhead for temperature Jacobians is only about a factor 2, see \citet{Schreier15}).
On the other hand, the time required for nonlinear least squares fitting (proportional to the number of iterations) can be reduced if a good initial guess is provided.
Of course the ``Chahine first guess'' can be used as initial guess for any other iterative optimisation solver.

In conclusion, we have used an extension of the classical Chahine approach in an exoplanet context for the first time to our knowledge.
This approach can deliver stable and reasonable temperature estimates for terrestrial-type exoplanets quickly (first guess in seconds, iterative refinements in minutes), even for challenging cases such as atmospheres with weak inversions and large \element{CO_2} abundances. 
It is also attractive in view of the growing awareness on the ``carbon footprint of large scale computing'' and green computing \citep[e.g.][]{Jahnke20}.
Hence it allows interesting new insight and provides a valuable addition to existing methods.

\section*{Acknowledgements}
This foundations for this research were the DFG projects SCHR 1125/3-1, RA-714/7-1, and RA 714/9-1. 
We acknowledge the support of the DFG priority programme SPP 1992 ``Exploring the Diversity of Extrasolar Planets (GO 2610/2-1)''.
J.L.G. thanks ISSI Team 464 for useful discussion.
We thank the German Research Foundation (DFG) for financial support via the project The Influence of Cosmic Rays on Exoplanetary Atmospheric Biosignatures (Project number 282759267).
Finally we would like to thank Joanna Barstow for the constructive review.

\section*{Data Availability}
Atmospheric data have been taken from \citet{Wunderlich19,Wunderlich20} and molecular spectroscopy data have been obtained from \textsc{Hitran} (\url{https://hitran.org/});
the data analysis software is an extension of the Py4CATS package \citep{Schreier19p} available at \url{https://atmos.eoc.dlr.de/tools/Py4CAtS/}.



\bibliographystyle{mnras}
\setlength{\bibsep}{0.5ex}



\bsp	
\label{lastpage}
\end{document}